\documentclass[preprint,aps]{revtex4}

\usepackage{amsmath}
\usepackage{color}
\usepackage{graphicx}
\usepackage{bm}
\usepackage{fancyhdr}
\usepackage{amsmath}
\usepackage{caption}
\usepackage{subcaption}
\usepackage{graphicx}
\usepackage{amsfonts}
\usepackage{amsbsy}
\usepackage{amssymb}
\usepackage[mathscr]{eucal}

\newcommand{\beq}{\begin{equation}}
\newcommand{\eeq}{\end{equation}}
\newcommand{\ba}{\begin{array}}
\newcommand{\ea}{\end{array}}
\newcommand{\bea}{\begin{eqnarray}}
\newcommand{\eea}{\end{eqnarray}}
\newcommand{\bseq}{\begin{subequations}}
\newcommand{\eseq}{\end{subequations}}

\begin{document}
	
\title{The influence of the symmetry of identical particles on flight times}
\author{Salvador Miret-Art\'es$^a$, Randall S. Dumont$^b$, Tom Rivlin$^c$ and Eli Pollak$^c$}
\affiliation{$^a$ Instituto de F\'isica Fundamental, CSIC, Serrano 123, 28006 Madrid, Spain}
\affiliation{$^b$ Department of Chemistry and Chemical Biology, McMaster University, Hamilton, Ontario, Canada L8S 4M1}
\affiliation{$^c$ Chemical and Biological Physics Department, Weizmann Institute of Science,
	76100 Rehovot, Israel}

\begin{abstract}
	
In this work, our purpose is to show how the symmetry of identical particles can influence  the time evolution of  free particles in the nonrelativistic and relativistic domains. For this goal, we consider a system of either two distinguishable or indistinguishable (bosons and fermions) particles. Two classes of initial conditions have been studied: different initial locations with the same momenta, and the same locations with different momenta. The flight time distribution of particles arriving at a `screen' is calculated in each case. Fermions display broader distributions as compared with either distinguishable particles or bosons, leading to earlier and later arrivals for all the cases analyzed here. The symmetry of the wave function seems to speed up or slow down propagation of particles. Due to the cross terms, certain initial conditions lead to bimodality in the fermionic case. 
Within the nonrelativistic domain and when the short-time survival probability is analyzed,  
if the cross term becomes important, one finds that the decay of the overlap of fermions 
is faster than for distinguishable particles which in turn is faster than for bosons. These results are of interest in the short time limit since they imply that the well-known quantum Zeno effect would be stronger for bosons than for fermions. 
Fermions also arrive earlier than bosons when they are scattered by a delta barrier. Furthermore, the particle symmetry does not affect the mean tunneling flight time and it is given by the phase time for the distinguishable particle.
	
\end{abstract}

\maketitle

\renewcommand{\theequation}{1.\arabic{equation}} \setcounter{section}{0} %
\setcounter{equation}{0}

\section{Introduction}

It is well understood that the symmetry of indistinguishable particles has a
profound influence on their dynamics. A feature which is well documented is
the ``bunching" of bosons \cite{brown1956,jeltes2007} and ``anti-bunching" of
fermions \cite{henny1999,oliver1999,kiesel2002,ianuzzi2006,rom2006}.
Consider two identical particles, each described for simplicity by an
initial Gaussian wavepacket. When the two Gaussians of the two particles are
located sufficiently far from each other in phase space, there is no overlap between them
and the symmetry of particles plays no role. The fermions and bosons may be
considered as two independent distinguishable particles. However, when they
come close the symmetry leads to important consequences. Bosons, whose overall function is
symmetric with respect to exchange may overlap with each other, hence the interference term
`increases' the density, causing the ``bunching'' phenomenon. Fermions
on the other hand, due to the anti-symmetry, cannot be located at the same
place and the `hole' in the distribution created by the overlap term creates
a distancing between the particles, which is understood as the
``anti-bunching" effect.

These effects show up also in the temporal dynamics \cite%
{grossmann2014,buchholz2018}. Consider the scattering of two
indistinguishable particles on each other and the relative distance
(squared) between them as a function of time \cite{grossmann2014}. As they
come closer to each other the distance is reduced and as they move again
away it increases. Yet, when comparing such scattering with the exact same
potential, incident energy, etc. of bosons and fermions, one finds that the
distance between the fermions as they separate is larger than that of bosons
-- another reflection of the bunching and anti-bunching phenomenon \cite%
{grossmann2014,buchholz2018}. Some researchers have tried to describe the
repulsion of fermions in terms of an artificial repulsive potential -- the ``Pauli potential". \cite{wilets1977,dorso1987,boal1988,latora1994,gu2016} In statistical mechanics, this situation leads to the so-called statistical interparticle potential which is temperature-dependent since 
it is related to what is known as the mean thermal wavelength or thermal de Broglie wavelength. \cite{Pathria} One then speaks about statistical 
attraction and repulsion for bosons and fermions, respectively. 

To the best of our knowledge the effect of symmetry on flight time
distributions \cite{petersen2017,petersen2018,rivlin2020,ianconescu2021} has
not been addressed. 
The central objective of this present work is to study how the symmetry affects temporal evolution, one-particle flight time distributions and, under the presence of an interaction potential, mean flight times when considering non-interacting identical particles. We will show that fermions have a
broader time distribution than bosons so that the former will be detected arriving at a suitably placed screen earlier and later, a direct result of the anti-bunching effect. 
Effectively, the symmetry can speed up or slow down the time evolving particles in the nonrelativistic and relativistic domains. In the second framework, we argue that one cannot speak of a superluminal effect since it can be seen as a mirror, or direct reflection, of the corresponding initial spatial distributions.

Specifically, consider first two identical particles initiated close to each other about a (mean) point in phase space, which then continue moving as free particles until they are detected on a screen some distance away. One may in principle measure the time at which a particle hits the screen
and thus obtain a flight time distribution. We will show that this flight time
distribution may be different for bosons and fermions as compared to
distinguishable particles. The same happens when their motion is not free but they
are individually scattered by a delta barrier potential.

We find  that fermions have a broader time distribution than bosons, irrespective of whether they are
scattered through a potential or not. Fermions will be detected arriving at
the screen earlier and later than bosons, a direct result of the anti-bunching effect.
A related question has to do with the survival probability of the initial
wavefunction. We shall show that the bosonic survival probability of free
particles decays slower than that of distinguishable particles while for fermions it decays more rapidly. These results are also of interest in the short time limit, which should imply that the well-known quantum Zeno effect \cite{Zeno1,Zeno2,Zeno3}
would be stronger for bosons than for fermions.
However, at least for scattering through a
delta potential, the particle symmetry does not affect the mean tunneling flight time and it is given by the phase time for the distinguishable particle \cite%
{rivlin2020,dumont2020}.

The paper is organized as follows. In Section II we consider the case of free particle propagation in the nonrelativistic and relativistic domains. In
Section III the scattering of identical particles from a delta barrier
potential is analyzed in detail since closed analytical expressions can be obtained.  Section IV presents and discusses our results for free and tunneling dynamics. The role played by the initial width of the Gaussian wavepackets describing the identical particles in mean flight times is also analyzed.
We argue that the implications of the early arrival of fermions versus bosons as, for example, photons at the screen in the relativistic case, which might seem to be superluminal, is not. It is a reflection of the anti-bunching effect on the initial density distribution.
We also consider further generalizations and
implications of these results to realistic systems in the last Section. 

\renewcommand{\theequation}{2.\arabic{equation}} \setcounter{section}{1} %
\setcounter{equation}{0}

\section{\protect\bigskip Free dynamics of nonrelativistic identical particles}

\subsection{General considerations}

Our model system is two (one dimensional) non-interacting 
identical particles (with coordinates $x_{1}$ and $x_{2})$ and mass$\ M\ $%
which scatter from a potential $V\left( x_{j}\right)$. We place a screen
to the right or left of the potential and measure a particle whenever it
hits the screen. The questions we seek to answer are what the
distribution of times at which one of the particles hits the screen is, and what
the mean time it takes is, assuming that the mean time exists. The
Hamiltonian for a single particle (operators are denoted with carets) is
\begin{equation}
\hat{H}_{j}=\frac{\hat{p}_{j}^{2}}{2M}+V\left( \hat{x}_{j}\right) ,j=1,2
\label{2.1}
\end{equation}%
with $\hat{p}_{j}$ and $\hat{x}_{j}$ the momentum and position operators of
the j-th particle respectively. The full Hamiltonian is the sum of the two%
\begin{equation}
\hat{H}=\hat{H}_{1}+\hat{H}_{2}.  \label{2.2}
\end{equation}%
Initially, the single particle wavefunction will be a coherent state
localized about the mean position $x_{ji}$ and mean momentum $p_{ji}$ with
width parameter $\Gamma$%
\begin{equation}
\Psi _{j}\left( x_{j}\right) =\left( \frac{\Gamma }{\pi }\right) ^{1/4}\exp %
\left[ -\frac{\Gamma \left( x_{j}-x_{ji}\right) ^{2}}{2}+\frac{i}{\hbar }%
p_{ji}\left( x_{j}-x_{ji}\right) \right] ,j=1,2.  \label{2.3}
\end{equation}%
To simplify, we introduce at this point the reduced coordinates of position, momenta, and time to be%
\begin{equation}
X=\sqrt{\Gamma }x,K=\frac{p}{\hbar \sqrt{\Gamma }},\tau =\frac{\hbar \Gamma
}{M}t  \label{2.4}
\end{equation}%
so that the single particle wavefunction has the form%
\begin{equation}
\Psi _{j}\left( X_{j}\right) =\left( \frac{1}{\pi }\right) ^{1/4}\exp \left[
-\frac{1}{2}\left( X_{j}-X_{ji}\right) ^{2}+iK_{ji}\left( X_{j}-X_{ji}\right) %
\right] ,j=1,2.  \label{2.5}
\end{equation}%
The composite wavefunction of the two particles is
\begin{equation}
\Psi _{k}\left( X_{1},X_{2}\right) =\frac{1}{N_{k}}\left[ \Psi _{1}\left(
X_{1}\right) \Psi _{2}\left( X_{2}\right) +h_{k}\Psi _{2}\left( X_{1}\right)
\Psi _{1}\left( X_{2}\right) \right]  \label{2.6}
\end{equation}%
where the coefficient $h_{k}$ is
\begin{equation}
h_{k}=\left(
\begin{array}{c}
1 \\
-1 \\
0%
\end{array}%
\right) \text{ for }\left(
\begin{array}{c}
\text{bosons} \\
\text{fermions} \\
\text{distinguishable particles}%
\end{array}%
\right)  \label{2.7}
\end{equation}%
and the corresponding normalization constant is
\begin{equation}
N_{k}^{2}=2\left[ 1+h_{k}\exp \left( -\frac{\left( X_{1i}-X_{2i}\right)
^{2}+\left( K_{1i}-K_{2i}\right) ^{2}}{2}\right) \right] \text{, }k=B,F
\label{2.8}
\end{equation}%
where $B$ and $F$ denote bosons and fermions, respectively. The
normalization for distinguishable particles ($k=D$) is unity. The time-evolved wavefunction is
\begin{eqnarray}
N_{k}\Psi _{k}\left( X_{1},X_{2};\tau \right) &=&\left[ \exp \left( -i\hat{H}%
_{1}\tau \right) \Psi _{1}\left( X_{1}\right) \right] \left[ \exp \left( -i%
\hat{H}_{2}\tau \right) \Psi _{2}\left( X_{2}\right) \right]  \notag \\
&&+h_{k}\left[ \exp \left( -i\hat{H}_{1}\tau \right) \Psi _{2}\left(
X_{1}\right) \right] \left[ \exp \left( -i\hat{H}_{2}\tau \right) \Psi
_{1}\left( X_{2}\right) \right]  \notag \\
&\equiv &\left[ \Psi _{1}\left( X_{1};\tau \right) \Psi _{2}\left(
X_{2};\tau \right) +h_{k}\Psi _{2}\left( X_{1};\tau \right) \Psi _{1}\left(
X_{2};\tau \right) \right] .  \label{2.9}
\end{eqnarray}

We now put a `screen' at the point $X=X_f$ such that the initial wave function has negligible overlap with the screen. 
Particle $1$ may reach the screen at
time $\tau $ \ when particle $2$ is found in any location. In other
words, the probability that a particle reaches the screen and that the second
particle will be found at that time at some point, say, $z$, will be
proportional to $\left\vert \Psi \left( X_f,X_{2}=z;\tau \right)
\right\vert ^{2}$. We are interested in knowing the distribution of times at
which a particle will hit the screen, irrespective of where the other
particle is so that the probability of finding a particle hitting the screen
at (the reduced) time $\tau $ is defined to be
\begin{equation}
P_{k}\left( X_f;\tau \right) =\frac{\int_{-\infty }^{\infty }dz\left\vert \Psi
_{k}\left( X_f,z;\tau \right) \right\vert ^{2}}{\int_{0}^{\infty }d\tau
\int_{-\infty }^{\infty }dz\left\vert \Psi _{k}\left( X_f,z;\tau \right)
\right\vert ^{2}},k=B,F,D.  \label{2.10}
\end{equation}%
The mean time is then naturally given as
\begin{equation}
\left\langle \tau \right\rangle _{k}=\int_{0}^{\infty }d\tau \tau
P_{k}\left( X_f;\tau \right) ,k=B,F,D.  \label{2.11}
\end{equation}%
The distribution and the means are well defined if the time integrals
converge as in potential scattering where the density decays at long times
as $\tau ^{-3}$ \cite{Muga-2008,Eli-2018}. For free particles, the density decays as $\tau ^{-1}$ so
that the best one can do is to consider the relative probability of having a
particle arrive at the screen at time $\tau $. This we denote as
\begin{equation}
\rho _{k}\left( X_f,\tau \right) =\int_{-\infty }^{\infty }dz\left\vert \Psi
_{k}\left( X_f,z;\tau \right) \right\vert ^{2},k=B,F,D.  \label{2.12}
\end{equation}

\subsection{Symmetry and free particles}

The single free particle ($V=0$) time evolved wavefunction is
\begin{equation}
\Psi _{j}\left( X,X_{i},K_{i},\tau \right) =\frac{1}{\sqrt{\left( 1+i\tau
\right) }}\left( \frac{1}{\pi }\right) ^{1/4}\exp \left( -\frac{1}{2}\frac{%
\left[ \left( X_{j}-X_{ji}\right) -iK_{ji}\right] ^{2}}{\left( 1+i\tau
\right) }-\frac{K_{ji}^{2}}{2}\right) ,j=1,2  \label{2.13}
\end{equation}%
and the density is
\begin{equation}
\left\vert \Psi _{j}\left( X,X_{ji},K_{ji},\tau \right) \right\vert ^{2}=%
\frac{1}{\sqrt{\pi \left( 1+\tau ^{2}\right) }}\exp \left( -\frac{\left(
X-X_{ji}-K_{ji}\tau \right) ^{2}}{\left( 1+\tau ^{2}\right) }\right) ,j=1,2.
\label{2.14}
\end{equation}%
The free particle time-dependent density has a maximum at the (free particle) time $\tau
=\left( X-X_{ji}\right) /K_{ji}$. It is normalized when integrating over the
position $X$ \ but diverges when integrating over the time due to the long
time tail which goes as $1/\tau $.

After some Gaussian integrations one finds that
\begin{eqnarray}
&&\rho _{k}\left( X;\tau \right) =\frac{1}{N_{k}^{2}}\left( \left\vert \Psi
_{1}\left( X,X_{1i},K_{1i},\tau \right) \right\vert ^{2}+\left\vert \Psi
_{2}\left( X,X_{2i},K_{2i},\tau \right) \right\vert ^{2}\right)  \notag \\
&&+h_{k}\frac{2}{N_{k}^{2}}\left\vert \Psi _{1}\left( X,X_{1i},K_{1i},\tau
\right) \right\vert \left\vert \Psi _{2}\left( X,X_{2i},K_{2i},\tau \right)
\right\vert  \notag \\
&&\exp \left( -\frac{\left( X_{2i}-X_{1i}\right) ^{2}+\left(
K_{2i}-K_{i1}\right) ^{2}}{4}\right) \cos \left( \Phi -\frac{\left(
X_{2i}-X_{1i}\right) \left( K_{2i}+K_{1i}\right) }{2}\right) ,k=B,F,D  \notag
\\
&&  \label{2.15}
\end{eqnarray}%
where the phase $\Phi $ is%
\begin{equation}
\Phi =\frac{\tau \left[ \left( X-X_{1i}\right) ^{2}-K_{1i}^{2}\right] }{%
2\left( 1+\tau ^{2}\right) }-\frac{\tau \left[ \left( X-X_{2i}\right)
^{2}-K_{2i}^{2}\right] }{2\left( 1+\tau ^{2}\right) }-\frac{K_{2i}\left(
X-X_{2i}\right) }{\left( 1+\tau ^{2}\right) }+\frac{\left( X-X_{1i}\right)
K_{1i}}{\left( 1+\tau ^{2}\right) }.  \label{2.16}
\end{equation}%
As also shown below, in the long-time limit ($\tau \rightarrow \infty $) the
density scales as $\rho _{k}\left( X;\tau \right) \sim 1/\tau $ irrespective of
whether one is considering bosons, fermions or distinguishable particles so
that strictly speaking for freely evolving particles the time integral in the denominator of Eq. \ref{2.10}
diverges.

\subsubsection{\protect\bigskip Bosons}

If the two bosons are initially placed such that $X_{1i}=X_{2i}=X_{i}$ and $%
K_{1i}=K_{2i}=K_{i}$ then the phase $\Phi $ vanishes, there is no effect of
interference, and the time dependent density is the same as for a
distinguishable particle. Similarly, if the initial distance between the two
wavepackets is sufficiently large, the interference cross term will vanish and the
result will again reduce to the single particle distinguishable case. The
interesting case is when the two particles are close to each other. If the
initial momenta are identical, that is, $K_{1i}=K_{2i}=K_{i}$ ($\Delta_{K}=K_{2i}-K_{1i}=0$) and the initial coordinates are written as average and difference coordinates
\begin{equation}
X_{i}=\frac{X_{1i}+X_{2i}}{2},\Delta _{X}=X_{2i}-X_{1i}  \label{2.17}
\end{equation}%
one finds that the density of finding a boson at the screen $X$\ at time $%
\tau $\
\begin{eqnarray}
&&\rho _{B}\left( X;\tau \right) \left( \Delta _{K}=0\right) =\frac{%
\left\vert \Psi \left( X,X_{i},K_{i},\tau \right) \right\vert ^{2}}{\left[
1+\exp \left( -\frac{\Delta _{X}^{2}}{2}\right) \right] }\exp \left( -\frac{%
\Delta _{X}^{2}}{4\left( 1+\tau ^{2}\right) }\right)  \notag \\
&&\left[ \cosh \left( \frac{\Delta _{X}\left( X-X_{i}-K_{i}\tau \right) }{%
\left( 1+\tau ^{2}\right) }\right) +\exp \left( -\frac{\Delta _{X}^{2}}{4}%
\right) \cos \left( \frac{\tau \Delta _{X}\left[ \left( X-X_{i}\right) -\tau
K_{i}\right] }{\left( 1+\tau ^{2}\right) }\right) \right]  \notag \\
&&  \label{2.18}
\end{eqnarray}%
and indeed one may check to see that%
\begin{equation}
\int_{-\infty }^{\infty }dX\rho _{B}\left( X;\tau \right) =1.  \label{2.19}
\end{equation}%
The long time limit is%
\begin{equation}
\lim_{\tau \rightarrow \infty }\rho _{B}\left( X;\tau \right) =\frac{%
\left\vert \Psi _{0}\left( X,X_{i},K_{i},\tau \right) \right\vert ^{2}}{%
\left[ 1+\exp \left( -\frac{\Delta _{X}^{2}}{2}\right) \right] }\left[
1+\exp \left( -\frac{\Delta _{X}^{2}}{4}\right) \cos \left( \Delta
_{X}K_{i}\right) \right]  \label{2.20}
\end{equation}%
so that as already mentioned, the free-particle time distribution of bosons
decays at long time as $\tau ^{-1}$ just like the single 
free particle.

Similarly let us consider two bosons that are initiated at the same point ($%
X_{1i}=X_{2i}=X_{i})$ but with different momenta and use the difference and
mean momenta%
\begin{equation}
K_{i}=\frac{K_{1i}+K_{2i}}{2},\Delta _{K}=K_{2i}-K_{1i} . \label{2.21}
\end{equation}%
In this case
\begin{eqnarray}
&&\rho _{B}\left( X;\tau \right) \left( \Delta _{X}=0\right) =\frac{2}{N^{2}%
\sqrt{\pi \left( 1+\tau ^{2}\right) }}\exp \left( -\frac{\left(
X-X_{i}-K_{i}\tau \right) ^{2}}{\left( 1+\tau ^{2}\right) }-\frac{\Delta
_{K}^{2}\tau ^{2}}{4\left( 1+\tau ^{2}\right) }\right)  \notag \\
&&\left[ \cosh \left( \frac{\Delta _{K}\tau \left( X-X_{i}-K_{i}\tau \right)
}{\left( 1+\tau ^{2}\right) }\right) +\exp \left( -\frac{\Delta _{K}^{2}}{4}%
\right) \cos \left( \frac{\Delta _{K}\left[ \tau K_{i}-\left( X-X_{i}\right) %
\right] }{\left( 1+\tau ^{2}\right) }\right) \right] .  \notag \\
&&  \label{2.23}
\end{eqnarray}%
and this again decays at long times as $\tau ^{-1}$.

\subsubsection{Fermions}

Following the same algebra as in the bosonic case, one finds that the
fermionic density is
\begin{eqnarray}
&&\rho _{F}\left( X;\tau \right) =\frac{1}{N_{F}^{2}}\left( \left\vert \Psi
\left( X,X_{1i},K_{1i},\tau \right) \right\vert ^{2}+\left\vert \Psi \left(
X,X_{2i},K_{2i},\tau \right) \right\vert ^{2}\right)  \notag \\
&&-\frac{2}{N_{F}^{2}}\left\vert \Psi \left( X,X_{1i},K_{1i},\tau \right)
\right\vert \left\vert \Psi \left( X,X_{2i},K_{2i},\tau \right) \right\vert
\notag \\
&&\cos \left( \Phi -\frac{\left( X_{2i}-X_{1i}\right) \left(
K_{2i}+K_{1i}\right) }{2}\right) \exp \left( -\frac{\left(
X_{2i}-X_{1i}\right) ^{2}+\left( K_{2i}-K_{1i}\right) ^{2}}{4}\right)
\label{2.24}
\end{eqnarray}%
The normalization is
\begin{equation}
N_{F}^{2}=2\left[ 1-\exp \left( -\frac{\Delta _{X}^{2}+\Delta _{K}^{2}}{2}%
\right) \right]   \label{2.25}
\end{equation}%
and it vanishes if $\Delta _{X}=\Delta _{K}=0$ so care must be taken in this
limit, since also the numerator vanishes but the ratio does not. More
specifically, at time $\tau =0$ we have with $K_{1i}=K_{2i}=K_{i}$ and $%
\Delta _{X}=X_{2i}-X_{1i}$
\begin{eqnarray}
&&\lim_{\Delta _{X}\rightarrow 0}\Psi \left( X_{1},X_{2};0\right) =-\frac{1}{%
\sqrt{\pi }}\left( X_{2}-X_{1}\right)  \notag \\
&&\cdot \exp \left[ iK_{i}\left( X_{1}-X_{1i}\right) +iK_{i}\left(
X_{2}-X_{2i}\right) -\frac{1}{2}\left[ \left( X_{1}-X_{1i}\right)
^{2}+\left( X_{2}-X_{2i}\right) ^{2}\right] \right]  \label{2.26}
\end{eqnarray}%
and this vanishes if $X_{1}=X_{2}$. Fermions cannot exist at the same point
in phase space. Note however that
\begin{equation}
\int_{-\infty }^{\infty }dX_{1}\int_{-\infty }^{\infty }dX_{2}\lim_{\Delta
_{i}\rightarrow 0}\left\vert \Psi \left( X_{1},X_{2};0\right) \right\vert
^{2}=1  \label{2.27}
\end{equation}%
as it should be. There is no difficulty in preparing an initial wavefunction
in the fermionic case even if both wavepackets are localized around the same
centers both in coordinate and momentum space. The density vanishes at one
point only.

Using the average and difference coordinates as above we readily find that
when the two incident momenta are identical
\begin{eqnarray}
&&\rho _{F}\left( X;\tau \right) \left( \Delta _{K}=0\right) =\frac{%
\left\vert \Psi _{0}\left( X,X_{i},K_{i},\tau \right) \right\vert ^{2}}{%
2\sinh \left( \frac{\Delta _{i}^{2}}{4}\right) }\exp \left( \frac{\tau
^{2}\Delta _{X}^{2}}{4\left( 1+\tau ^{2}\right) }\right)  \notag \\
&&\left[ \cosh \left( \frac{\Delta _{i}\left( X-X_{i}-K_{i}\tau \right) }{%
\left( 1+\tau ^{2}\right) }\right) -\exp \left( -\frac{\Delta _{X}^{2}}{4}%
\right) \cos \left( \frac{\tau \Delta _{X}\left[ \left( X-X_{i}\right) -\tau
K_{i}\right] }{\left( 1+\tau ^{2}\right) }\right) \right] . \notag \\
&&  \label{2.28}
\end{eqnarray}%
It is also straightforward to see that
\begin{equation}
\int_{-\infty }^{\infty }dX\rho _{F}\left( X;\tau \right) \left( \Delta
_{K}=0\right) =1.  \label{2.29}
\end{equation}%
When the (mean) distance between the two particles becomes small
\begin{equation}
\lim_{\Delta _{X}\rightarrow 0}\rho _{F}\left( X;\tau \right) =\left\vert
\Psi _{0}\left( X,X_{i},K_{i},\tau \right) \right\vert ^{2}\left[ \frac{%
\left( X-X_{i}-K_{i}\tau \right) ^{2}}{\left( 1+\tau ^{2}\right) }+\frac{1}{2%
}\right]  \label{2.30}
\end{equation}%
and at long times the fermion density is\bigskip
\begin{equation}
\lim_{\tau \rightarrow \infty }\rho _{F}\left( X;\tau \right) \left( \Delta
_{K}=0\right) =\frac{\left\vert \Psi _{0}\left( X,X_{i},K_{i},\tau \right)
\right\vert ^{2}}{\left[ 1-\exp \left( -\frac{\Delta _{X}^{2}}{2}\right) %
\right] }\left[ 1-\exp \left( -\frac{\Delta _{i}^{2}}{4}\right) \cos \left(
K_{i}\Delta _{X}\right) \right]  \label{2.31}
\end{equation}%
and it too decays as $\tau ^{-1}$ as for bosons. The difference is only in
the coefficients.

\subsection{Survival probability and symmetry}

The time-dependent overlap or survival amplitude for a single particle is%
\begin{eqnarray}
S_{j}\left( \tau \right) &=&\langle \Psi \left( X_{ji},K_{ji},0\right) |\Psi
\left( X_{ji},K_{ji},\tau \right) \rangle  \notag \\
&=&\sqrt{\frac{2}{\left( 2+i\tau \right) }}\exp \left( -\frac{i\tau
K_{ji}^{2}}{\left( 2+i\tau \right) }\right) . \label{2.32}
\end{eqnarray}%
Analogously, the time-dependent overlap or survival amplitude for the two particle wavefunction is then
\begin{equation*}
\Sigma \left( \tau \right) =\frac{2}{N_{k}^{2}}S_{1}\left( \tau \right)
S_{2}\left( \tau \right) \left[ 1+h_{k}\exp \left( -\frac{\Delta
_{X}^{2}+\Delta _{K}^{2}}{\left( 2+i\tau \right) }\right) \right]
\end{equation*}%
and its square is%
\begin{equation}
\left\vert \Sigma _{k}\left( \tau \right) \right\vert ^{2}=\left\vert
S_{1}\left( \tau \right) S_{2}\left( \tau \right) \right\vert
^{2}O_{k}\left( \tau \right)  \label{2.33}
\end{equation}%
with
\begin{eqnarray}
O_{k}\left( \tau \right) & =&\frac{\left[ 1+h_{k}^{2}\exp \left( -\frac{%
4\left( \Delta _{X}^{2}+\Delta _{K}^{2}\right) }{\left( 4+\tau ^{2}\right) }%
\right) +2h_{k}\exp \left( -\frac{2\left( \Delta _{X}^{2}+\Delta
_{K}^{2}\right) }{\left( 4+\tau ^{2}\right) }\right) \cos \left( \frac{%
\left( \Delta _{X}^{2}+\Delta _{K}^{2}\right) }{\left( 4+\tau ^{2}\right) }%
\tau \right) \right] }{\left[ 1+h_{k}\exp \left( -\frac{\Delta
_{X}^{2}+\Delta _{K}^{2}}{2}\right) \right] ^{2}},  \notag \\
k&=&B,F,D.  \label{2.34}
\end{eqnarray}

It is then of interest to study this overlap $ O_{k}\left( \tau \right)$ in some limits. First, we note
that when the initial distances between the wavepackets are sufficiently
large, such that $\Delta _{X}^{2}+\Delta _{K}^{2}\gg 1$, then for times
shorter than $\sim \sqrt{\Delta _{X}^{2}+\Delta _{K}^{2}}$, this overlap
function reduces to unity. This is what is expected: when the initial
distance between the particles is large, they behave as independent
distinguishable particles. The interesting case is when the initial
distances between the two wavepackets are small and the interference term is
no longer negligible at short times. For bosons, one finds to leading order
\begin{equation}
\lim_{\Delta _{X},\Delta _{K}\rightarrow 0}O_{B}\left( \tau \right) =1+\frac{%
\left( \Delta _{X}^{2}+\Delta _{K}^{2}\right)
\tau ^{2}}{2\left( 4+\tau
^{2}\right) }\geq 1 =O_{D}\left( \tau \right)   \label{2.35}
\end{equation}%
showing that in this limit, the bosonic survival probability is greater than
the distinguishable particle overlap and this is so for all times. For
fermions, though, one has that
\begin{equation}
\lim_{\Delta _{X},\Delta _{K}\rightarrow 0}O_{F}\left( \tau \right) =\frac{4%
}{\left( 4+\tau ^{2}\right) }\leq 1=O_{D}\left( \tau \right)   \label{2.37}
\end{equation}%
showing that the fermionic overlap decays faster than the distinguishable
particle case in this limit. In other words, when the distance in phase
space between the centers of the two particles is small, which is the case
when the interference term becomes most important, one finds that the decay
of the overlap of fermions is faster than distinguishable particles which in
turn is faster than bosons. These results are also of interest in the short
time limit, where one finds that%
\begin{equation}
\lim_{\Delta _{X},\Delta _{K},\tau \rightarrow 0}O_{B}\left( \tau \right) =1+%
\frac{\left( \Delta _{X}^{2}+\Delta _{K}^{2}\right) \tau ^{2}}{8}
\label{2.38}
\end{equation}%
\begin{equation}
\lim_{\Delta _{X},\Delta _{K},\tau \rightarrow 0}O_{F}\left( \tau \right)
=\left( 1-\frac{\tau ^{2}}{4}\right)   \label{2.39}
\end{equation}%
which indicates that the quantum Zeno effect \cite{Zeno1,Zeno2,Zeno3} would be stronger for bosons
than for fermions, as the fermionic survival probability decays faster also
at short times.

\subsection{\protect\bigskip Free dynamics of relativistic identical particles}

To investigate the relativistic regime, we consider relativistic electrons and photons.  The wavepackets describing the bosons -- the photons -- travel dispersion-free at the speed of light.  The wavepackets describing the fermions -- the electrons -- are four component spinors with time evolution determined by the Dirac equation.  As we consider only free particle motion of two non-interacting (except via particle statistics) electrons, spin is conserved and the wavepackets reduce to two component spinors.  Relativistic wavepacket propagation is much like non-relativistic propagation except that the velocity is no longer directly proportional to wavenumber -- the former asymptotes to the speed of light -- and wavepacket broadening is greatly suppressed due to the dispersion relation -- quadratic in the non-relativistic case -- approaching linearity.  In particular, the time scale for wavepacket broadening scales with $\gamma^2$, where  $\gamma = 1/\sqrt{1-v^2/c^2}$. (See Eq. (2.19) in \cite{dumont2020}.)  For example, if $v=0.99c$, broadening takes 50 times longer than it does for non-relativistic velocities.

The single free relativistic electron time evolved wavefunction, in the highly accurate (for cases we considered) steepest descent approximation, is
\begin{equation}
\Psi _{j}\left( X,X_{i},K_{i},\tau \right) =\frac{\hat{u}}{\sqrt{\left( 1+i\left(\tau/\gamma^2 \right)
\right) }}\left( \frac{1}{\pi }\right) ^{1/4}\exp \left( -\frac{1}{2}\frac{%
\left[ \left( X_{j}-X_{ji}\right) -iK_{ji}\right] ^{2}}{\left( 1+i\left(\tau/\gamma^2 \right)
\right) }-\frac{K_{ji}^{2}}{2}\right) ,j=1,2,  \label{3.1}
\end{equation}%
where $\hat{u}=u/\lVert u \rVert$ and
\begin{equation}
u=\left(
\begin{array}{c}
1 \\
\left(\frac{\hbar \Gamma^{1/2}}{mc} \right) \frac{K_{ji}}{1+\gamma}
\end{array}%
\right)
\end{equation}
is the two component spinor for a spin up electron. This is then used in the symmetrized wavefunction for the two bosons and electrons, respectively. 

\bigskip 

\bigskip

\renewcommand{\theequation}{4.\arabic{equation}} \setcounter{section}{3} %
\setcounter{equation}{0}

\section{\protect\bigskip Flight times of identical non-relativistic particles scattered by a
delta function barrier}

\subsection{Preliminaries}

The Hamiltonian for the delta function barrier is
\begin{equation}
\hat{H}=-\frac{\hbar ^{2}}{2M}\frac{d^{2}}{dx^{2}}+\varepsilon \delta \left(
x\right) .  \label{3.1}
\end{equation}%
and the coupling coefficient $\varepsilon >0$. The eigenfunctions of the
Hamiltonian at energy
\begin{equation}
E=\frac{\hbar ^{2}k^{2}}{2M}  \label{3.2}
\end{equation}%
are
\begin{equation}
\psi \left( x\right) =\left(
\begin{array}{c}
\exp \left( ikx\right) +R\left( k\right) \exp \left( -ikx\right) ,\text{ \ \
}x<0 \\
T\left( k\right) \exp \left( ikx\right) ,\text{ \ \ \ \ \ \ \ \ \ \ \ \ \ \
\ \ \ \ \ \ }x>0%
\end{array}%
\right)  \label{3.3}
\end{equation}%
with the reflection amplitude given as
\begin{equation}
R\left( k\right) =\frac{-i\alpha \left( k\right) }{1+i\alpha \left( k\right)
},\text{ \ \ }\alpha \left( k\right) =\frac{M\varepsilon }{\hbar ^{2}k}.
\label{3.4}
\end{equation}%
The transmission amplitude is
\begin{equation}
T\left( k\right) =\frac{1}{1+i\alpha \left( k\right) }  \label{3.5}
\end{equation}%
and one readily sees that
\begin{equation}
\left\vert R\left( k\right) \right\vert ^{2}+\left\vert T\left( k\right)
\right\vert ^{2}=1.  \label{3.6}
\end{equation}

The phase time delays are defined to be
\begin{equation}
\delta t_{T,R}=\frac{M}{\hbar k}\mathrm{Im}\left( \frac{1}{Y}\frac{\partial Y%
}{\partial k}\right) ,\text{ \ \ }Y=R,T  \label{3.7}
\end{equation}%
so that
\begin{equation}
\delta t_{T}=\delta t_{R}=\frac{M}{\hbar k}\frac{\alpha \left( k\right) }{k%
\left[ 1+\alpha ^{2}\left( k\right) \right] }\equiv \delta t.  \label{3.8}
\end{equation}%
The phase time then implies that for a repulsive delta function potential ($%
\alpha \left( k\right) >0$) the flight time is lengthened while for an
attractive delta function potential it is shortened. In the limit that the
coupling coefficient $\varepsilon \rightarrow \infty $, which is the
equivalent of a hard wall potential, the transmission amplitude vanishes
while the reflection amplitude goes to $-1$. The reflection time delay
vanishes in this case, the interference of the forward and reflected wave
does not change the reflected phase time delay. For a fixed nonzero value
of $\varepsilon >0$ in the limit that the energy vanishes ($k\rightarrow 0)$
the delay diverges as $k^{-1}$. For an attractive delta function potential,
the flight time is shortened and the reduction diverges as $k^{-1}$. Due to
the zero width of the delta function potential, the dwell time \cite{Mugabook} in the
barrier always vanishes.

It is worthwhile here also to consider the imaginary time defined as \cite{Pollak1984}
\begin{equation}
t_{im,R,T}=\hbar \text{Re}\left( \frac{1}{Y}\frac{\partial Y}{\partial E}%
\right) ,\text{ \ \ }Y=R,T  \label{3.9}
\end{equation}%
so that the transmitted imaginary time is positive%
\begin{equation}
t_{im,T}=\frac{M\alpha ^{2}\left( k\right) }{\hbar k^{2}\left( 1+\alpha
^{2}\left( k\right) \right) }  \label{3.10}
\end{equation}%
while the reflected imaginary time is negative%
\begin{equation}
t_{im,R}=-\frac{M}{\hbar k^{2}\left( 1+\alpha ^{2}\left( k\right) \right) }.
\label{3.11}
\end{equation}%
As the momentum increases, the transmission probability increases while the
reflection probability decreases, so that the transmitted imaginary time is
positive, and the reflected is negative.

In reduced variables (with $\epsilon_i=\alpha(k)$), the phase time delay takes the simple form%
\begin{equation}
	\delta \tau \left( K_{i}\right) =\frac{\epsilon _{i}}{K_{i}^{2}\left(
		1+\epsilon _{i}^{2}\right) }
\end{equation}%
while the imaginary time delays are
\begin{equation}
	\tau _{im,T}\left( K_{i}\right) =\frac{\epsilon _{i}^{2}}{K_{i}^{2}\left(
		1+\epsilon _{i}^{2}\right) }
\end{equation}%
and%
\begin{equation}
	\tau _{im,R}\left( K_{i}\right) =-\frac{1}{K_{i}^{2}\left( 1+\epsilon
		_{i}^{2}\right) }.
\end{equation}%
The transmission and reflection probabilities become
\begin{equation}
	\left\vert T\left( K_{i}\right) \right\vert ^{2}=\frac{1}{1+\epsilon _{i}^{2}%
	},\left\vert R\left( K_{i}\right) \right\vert ^{2}=\frac{\epsilon _{i}^{2}}{%
		1+\epsilon _{i}^{2}}.
\end{equation}%

\subsubsection{The single particle dynamics. Momentum filtering}

Initially we consider a Gaussian wavepacket as in Eq. \ref{2.5}
\begin{equation}
\Psi \left( x,0\right) =\left( \frac{\Gamma }{\pi }\right) ^{1/4}\exp \left[
-\frac{\Gamma }{2}\left( x-x_{i}\right) ^{2}+ik_{i}\left( x-x_{i}\right) %
\right]  \label{3.12}
\end{equation}%
whose momentum representation is
\begin{equation}
\Psi \left( k,0\right) =\left( \frac{1}{\pi \Gamma }\right) ^{1/4}\exp
\left( -\frac{\left( k-k_{i}\right) ^{2}}{2\Gamma }-ikx_{i}\right) .
\label{3.13}
\end{equation}%
The time dependent wavepacket in the transmitted region is
\begin{equation}
\Psi _{T}\left( x,t\right) =\int_{-\infty }^{\infty }\frac{dk}{\sqrt{2\pi }}%
\Psi \left( k,0\right) T\left( k\right) \exp \left( ikx-i\frac{\hbar k^{2}}{%
2M}t\right) ,x\geq 0  \label{3.14}
\end{equation}%
and in the reflected region is%
\begin{equation}
\Psi _{R}\left( x,t\right) =\int_{-\infty }^{\infty }\frac{dk}{\sqrt{2\pi }}%
\exp \left( -i\frac{\hbar k^{2}}{2M}t\right) \Psi \left( k,0\right) \left[
\exp \left( ikx\right) +R\left( k\right) \exp \left( -ikx\right) \right]
,x\leq 0  \label{3.15}
\end{equation}

Using the reduced variables as in Eq. \ref{2.4} and the reduced delta
function coupling variable
\begin{equation}
\epsilon =\frac{M\varepsilon }{\hbar ^{2}\sqrt{\Gamma }}  \label{3.16}
\end{equation}%
and carrying out the momentum integrations in Eqs. \ref{3.14} and \ref{3.15}
one finds that the transmitted time-dependent wavepacket ($X\geq 0$) is
\begin{eqnarray}
&&\Psi _{T}\left( X,\tau \right) =\Psi _{fp}\left( X,\tau \right)  \notag \\
&&\left[ 1-\epsilon \frac{\sqrt{\pi \left( 1+i\tau \right) }}{\sqrt{2}}\exp %
\left[ -\frac{\left( 1+i\tau \right) }{2}\left( Z_{T}+i\epsilon \right) ^{2}%
\right] \mathrm{{erf}c}\left( -\frac{i\sqrt{\left( 1+i\tau \right) }}{\sqrt{2%
}}\left( Z_{T}+i\epsilon \right) \right) \right] .  \label{3.17}
\end{eqnarray}%
where $\Psi _{fp}\left( X,\tau \right) $ is the free particle time-dependent
wavepacket as in Eq. \ref{2.13}, $\mathrm{erfc}$ is the complementary
error function, and
\begin{equation}
Z_{T}=\frac{\left[ K_{i}+i\left( X-X_{i}\right) \right] }{\left( 1+i\tau
\right) }.  \label{3.18}
\end{equation}%
The reflected time-dependent wavefunction ($X\leq 0$) is
\begin{eqnarray}
&&\Psi _{R}\left( X,\tau \right) =\Psi _{fp}\left( X,\tau \right)  \notag \\
&&-\Psi _{fp}\left( -X,\tau \right) \epsilon \frac{\sqrt{\pi }}{\sqrt{2}}%
\sqrt{\left( 1+i\tau \right) }\exp \left( -\frac{\left( 1+i\tau \right) }{2}%
\left( i\epsilon +Z_{R}\right) ^{2}\right) \mathrm{\mathrm{{erf}c}}\left( -i%
\sqrt{\frac{\left( 1+i\tau \right) }{2}}\left( i\epsilon +Z_{R}\right)
\right)  \notag \\
&&  \label{3.19}
\end{eqnarray}%
with
\begin{equation}
Z_{R}=\frac{\left[ K_{i}-i\left( X+X_{i}\right) \right] }{\left( 1+i\tau
\right) }.  \label{3.20}
\end{equation}

In practice, if the incident (reduced) momentum is sufficiently large, which
will be the case in all of our computations, and since we will be using small
momentum variances, one may safely replace the complementary error function
with its asymptotic expansion so that to leading order
\begin{equation}
\Psi _{T}\left( X,\tau \right) \simeq \Psi _{fp}\left( X,\tau \right) \left[
\frac{Z_{T}}{\left( Z_{T}+i\epsilon \right) }\right] ,X\geq 0  \label{3.21}
\end{equation}%
and
\begin{equation}
\Psi _{R}\left( X,\tau \right) \simeq \Psi _{fp}\left( X,\tau \right) -\Psi
_{fp}\left( -X,\tau \right) \frac{\epsilon }{\left( \epsilon -iZ_{R}\right) }%
,X\leq 0.  \label{3.22}
\end{equation}%
Eqs. \ref{3.21} and \ref{3.22} will be the `workhorses' for the numerical
implementations below, but we stress that we have checked the validity of the
asymptotic expansion and it is quantitative for the conditions here used. To
see the long time limit we note that
\begin{equation}
\left\vert \frac{Z_{T}}{\left( Z_{T}+i\epsilon \right) }\right\vert ^{2}=%
\frac{\left[ K_{i}^{2}+\left( X-X_{i}\right) ^{2}\right] }{\left[ \left(
K_{i}-\epsilon \tau \right) ^{2}+\left( X-X_{i}+\epsilon \right) ^{2}\right]
}  \label{3.23}
\end{equation}
and this goes as $\tau ^{-2}$ in the long time limit so that the transmitted
single particle density decays as $\tau ^{-3}$. The calculation is a bit
more involved for the reflected density but it also decays as $\tau ^{-3}$.
In contrast to the free particle, due to the potential, the mean flight time
(Eqs. \ref{2.10} and \ref{2.11}) is well-defined.

We can now rewrite the density as
\begin{equation}
	\left\vert \Psi _{T}\left( X,\tau \right) \right\vert ^{2}\simeq \frac{%
		\left( 1+\tau _{fp}^{2}\right) }{\sqrt{\pi \left( 1+\tau ^{2}\right) }}\exp
	\left( -\frac{K_{i}^{2}\left[ \tau _{fp}-\tau \right] ^{2}}{\left( 1+\tau
		^{2}\right) }-\ln \left[ \left( \epsilon _{i}\tau -1\right) ^{2}+\left( \tau
	_{fp}+\epsilon _{i}\right) ^{2}\right] \right)
\end{equation}%
and ask when the exponent is maximized as a function of the reduced time $%
\tau $. Defining
\begin{equation}
	G\left( \tau \right) =\frac{K_{i}^{2}\left[ \tau _{fp}-\tau \right] ^{2}}{%
		\left( 1+\tau ^{2}\right) }+\ln \left[ \left( \epsilon _{i}\tau -1\right)
	^{2}+\left( \tau _{fp}+\epsilon _{i}\right) ^{2}\right]
\end{equation}%
we note that%
\begin{equation}
	\frac{dG\left( \tau \right) }{d\tau }=-2\frac{K_{i}^{2}\left[ \tau
		_{fp}-\tau \right] }{\left( 1+\tau ^{2}\right) }-2\tau \frac{K_{i}^{2}\left[
		\tau _{fp}-\tau \right] ^{2}}{\left( 1+\tau ^{2}\right) ^{2}}+\frac{\epsilon
		_{i}2\left( \epsilon _{i}\tau -1\right) }{\left[ \left( \epsilon _{i}\tau
		-1\right) ^{2}+\left( \tau _{fp}+\epsilon _{i}\right) ^{2}\right] }.
\end{equation}%
Setting the derivative equal to zero
\begin{equation}
	\frac{K_{i}^{2}\left[ \tau _{fp}-\tau \right] }{\left( 1+\tau ^{2}\right) }%
	+\tau \frac{K_{i}^{2}\left[ \tau _{fp}-\tau \right] ^{2}}{\left( 1+\tau
		^{2}\right) ^{2}}=\frac{\epsilon _{i}\left( \epsilon _{i}\tau -1\right) }{%
		\left[ \left( \epsilon _{i}\tau -1\right) ^{2}+\left( \tau _{fp}+\epsilon
		_{i}\right) ^{2}\right] }.
\end{equation}%
and looking for a solution
\begin{equation}
	\tau =\tau _{fp}\left( 1-\Delta \tau \right)
\end{equation}%
and assuming that $\Delta \tau \ll 1$ and remembering that $\tau _{fp}\gg 1$
leads to the solution%
\begin{equation}
	\Delta \tau \simeq \frac{\epsilon _{i}^{2}}{K_{i}^{2}\left[ 1+\epsilon
		_{i}^{2}\right] }=\tau _{im,T}\left( K_{i}\right)
\end{equation}%
and this is precisely the momentum filtering effect. Due to the increase of the transmission probability with energy, the high-energy components of the incident wavepacket are preferably transmitted so that the flight time is reduced \cite{Filinov}.

\subsubsection{Two particles dynamics}

The composite initial wavefunction of the two particles is given by Eq. (\ref{2.6}) and the time evolved wavefunctions by Eq. (\ref{2.9}) for free particle evolution. For the $\delta$-tunneling dynamics, the initial wave function is the same but Eq. (\ref{2.9}) has to be replaced by 
\begin{equation}\label{3.34}
\Psi _{k,T}\left( X, z, \tau \right) = \Psi _{1,T}\left( X,\tau \right) \left[\Psi _{2,T}\left(z,\tau \right)+\Psi _{2,R}\left(z,\tau \right)\right] + h_k \, \Psi _{2,T}\left( X,\tau \right)   \left[\Psi _{1,T}\left(z,\tau \right)+\Psi _{1,R}\left(z,\tau \right)\right]  \, 
\end{equation}
for the total transmitted wavefunction and
\begin{equation}\label{3.35}
	\Psi _{k,R}\left( X, z, \tau \right) = \Psi _{1,R}\left( X,\tau \right) \left[\Psi _{2,T}\left(z,\tau \right)+\Psi _{2,R}\left(z,\tau \right)\right] + h_k \, \Psi _{2,R}\left( X,\tau \right)   \left[\Psi _{1,T}\left(z,\tau \right)+\Psi _{1,R}\left(z,\tau \right)\right] \, 
\end{equation}
for the total reflected wavefunction, $ \Psi _{1,T}$, $ \Psi _{2,T}$, $ \Psi _{1,R}$ and $ \Psi _{2,R}$ being the 
transmitted and reflected wave functions for each particle when considered to be independent. The one-particle mean flight time is given by 
\begin{equation}
\left\langle \tau_k \right\rangle _{T,R}=\int_{0}^{\infty }d\tau \, \tau \, P_{k;T,R}\left( X_f, \tau \right) ,   \label{3.36}
\end{equation}%
where the probability distribution is
\begin{equation}
	P_{k;T,R}\left( X_f, \tau \right) =\frac{\int_{-\infty }^{\infty }dz\left\vert \Psi
		_{k;T,R}\left( X_f,z;\tau \right) \right\vert ^{2}}{\int_{0}^{\infty }d\tau
		\int_{-\infty }^{\infty }dz\left\vert \Psi _{k;T,R}\left( X_f,z,\tau \right)
		\right\vert ^{2}}  \label{3.37}
\end{equation}%
with $\, k=D, B, F$. The screen is located at $X=\pm X_f$ depending on whether we are considering the transmitted (plus sign) or reflected (minus sign) total wave function.  As mentioned above,
these distributions and mean times are well defined under the presence of a potential since the density decays at long times
as $\tau ^{-3}$.

\renewcommand{\theequation}{5.\arabic{equation}} \setcounter{section}{4} %
\setcounter{equation}{0}
\section{Numerical Results}

\subsection{Free particle non-relativistic flight times}

To analyze the role played by the symmetry of the total wave function in systems of identical particles when considering free dynamics 
and tunneling from a delta barrier, we have identified two types of initial conditions (in reduced coordinates): (I) different locations with the same 
momenta, $X_{1i}=-301$, $X_{2i}= -299$ with $K_{1i}=K_{2i}=10$  and (II) the same locations with different momenta, $X_{1i}=X_{2i}= -300$ 
with $K_{1i}=10.1$ and $K_{2i}=9.9$. When the differences in initial positions 
and momenta differ more, the cross terms in the density of the two-particle system become smaller, and one rapidly reaches the 
distinguishable particle limit. Unless otherwise stated, we will assume $\Gamma = 0.01$ for the initial width of the Gaussian functions and
$\epsilon = 1$, for the strength of the coupling to the delta barrier. In all cases, the position of the screen is at $X_f= \pm 450$ and the delta barrier is located
at the origin.

\begin{figure}
    \centering
    \begin{subfigure}[t]{0.45\textwidth}
        \centering	
        \includegraphics[width=\textwidth]{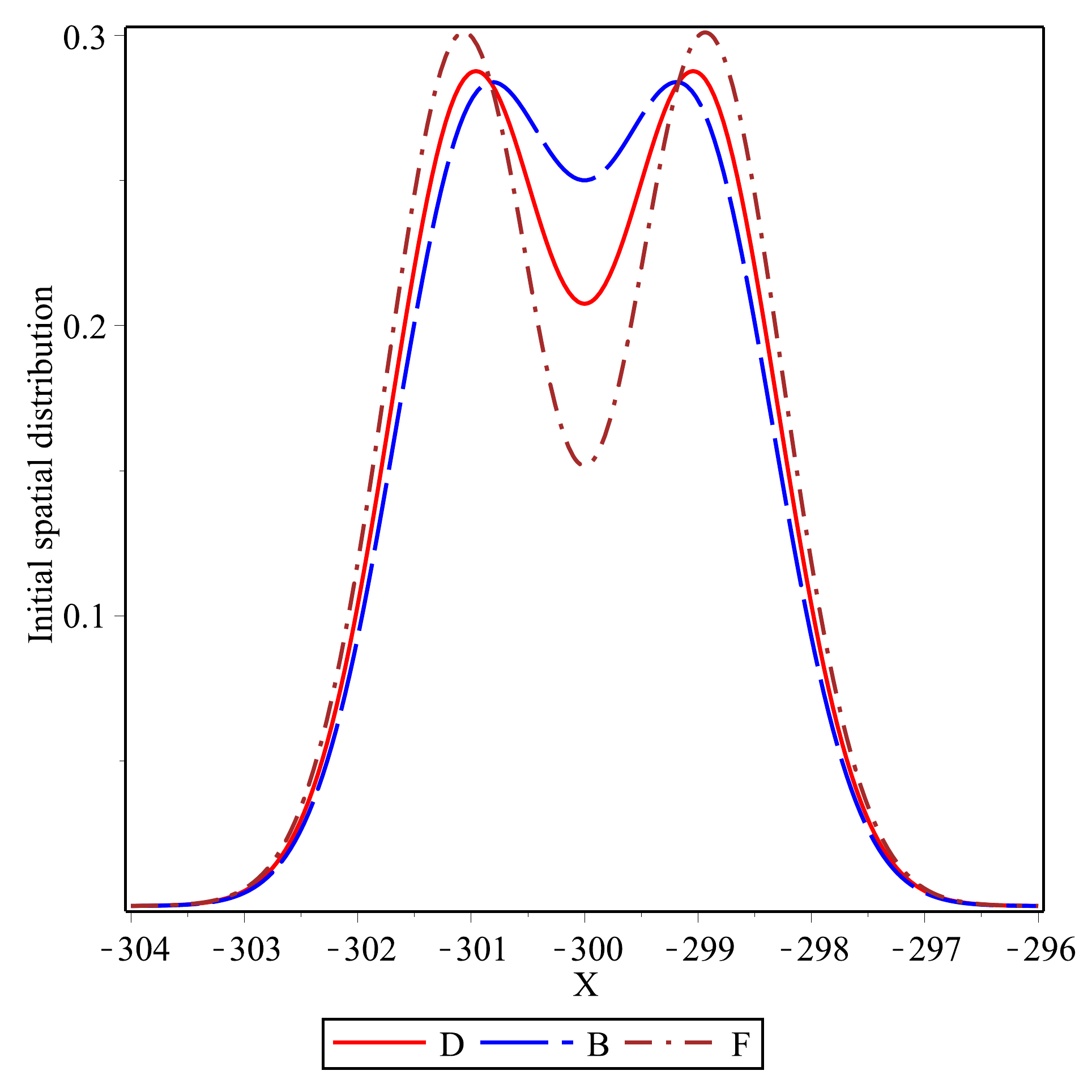}
        \label{fig1a}
    \end{subfigure}
\hfill
    \begin{subfigure}[t]{0.45\textwidth}
        \centering	
        \includegraphics[width=\textwidth]{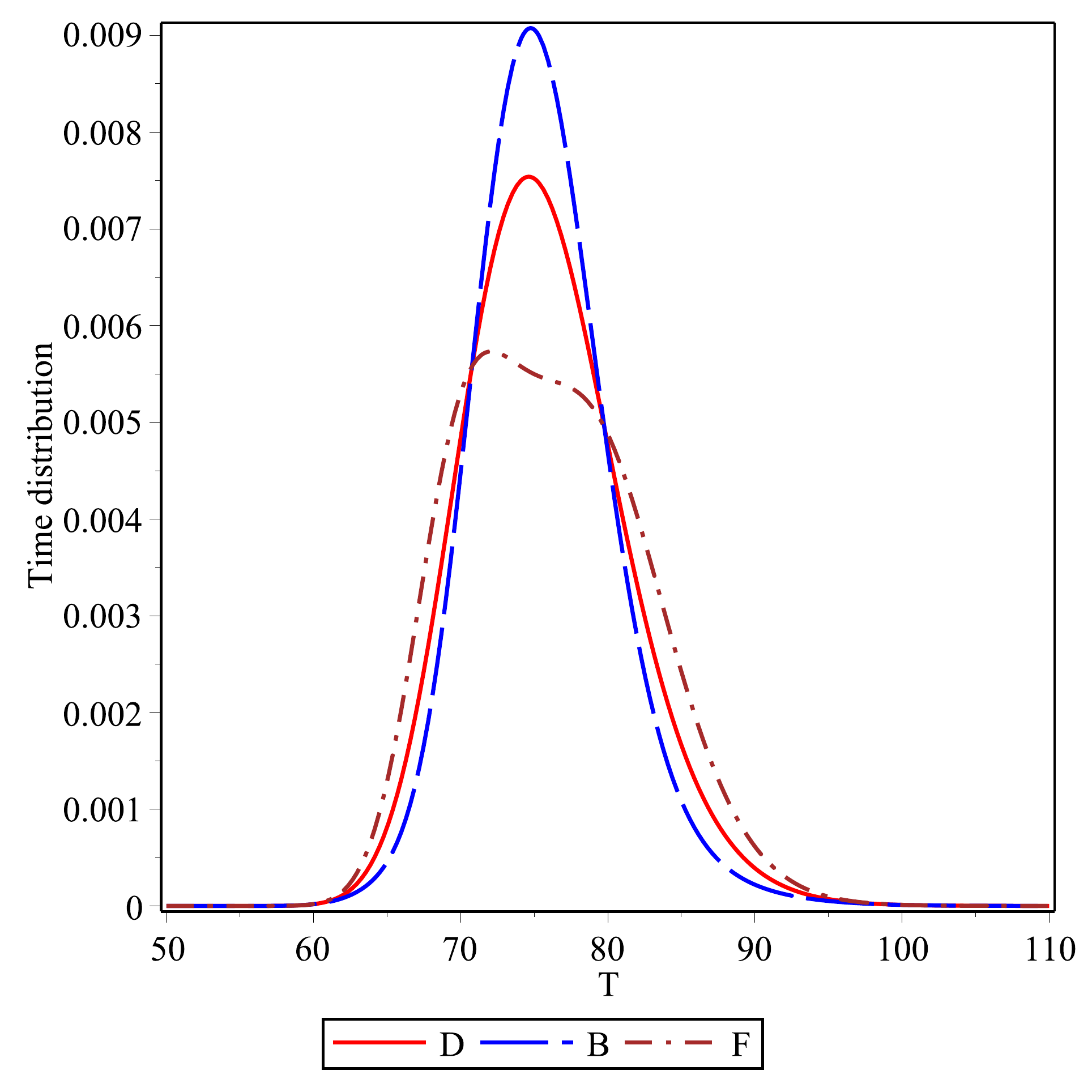}
        \label{fig1b}
    \end{subfigure}
    \begin{subfigure}[b]{0.45\textwidth}
	    \centering	
    	\includegraphics[width=\textwidth]{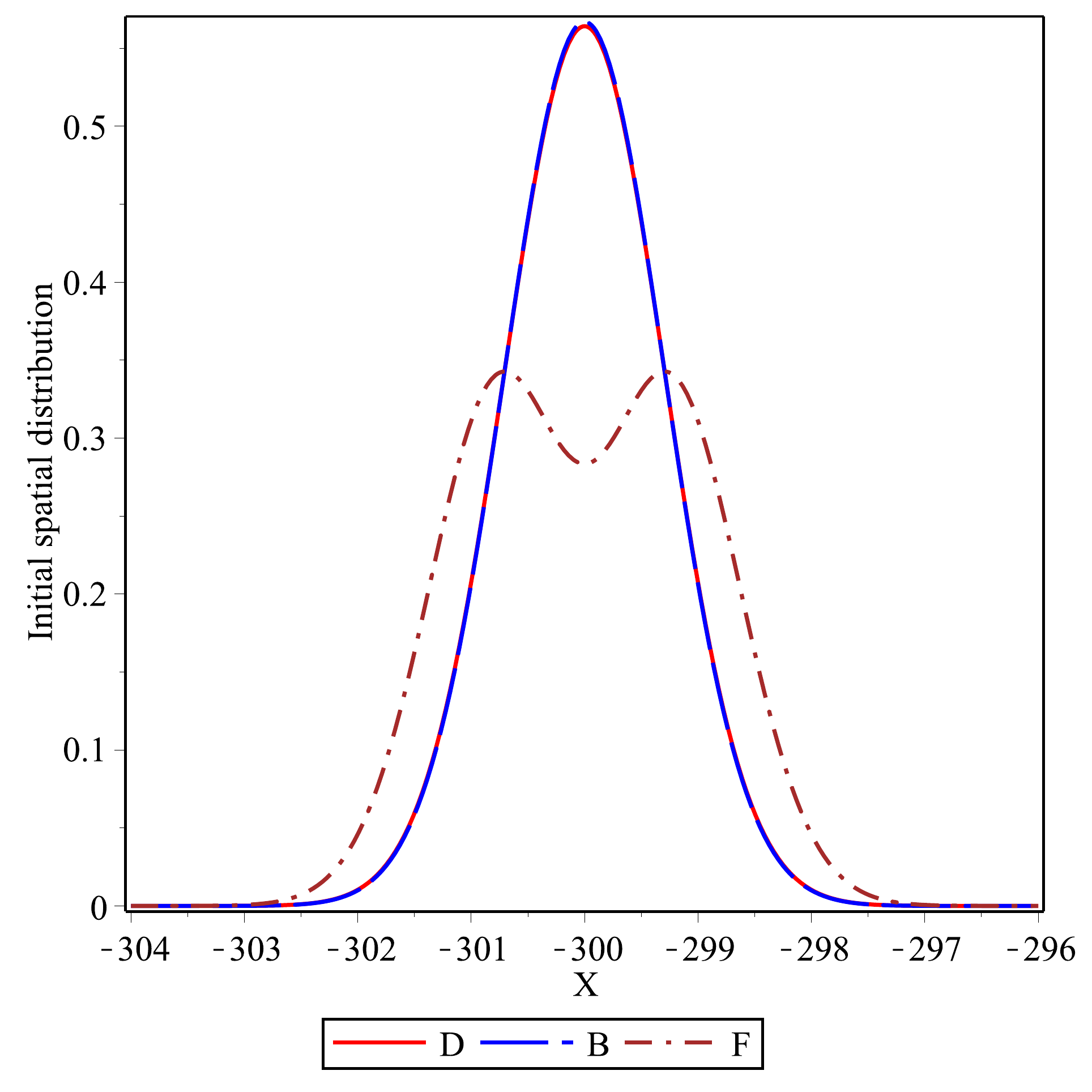}
	    \label{fig1c}
    \end{subfigure}
\hfill
    \begin{subfigure}[b]{0.45\textwidth}
    	\centering	
    	\includegraphics[width=\textwidth]{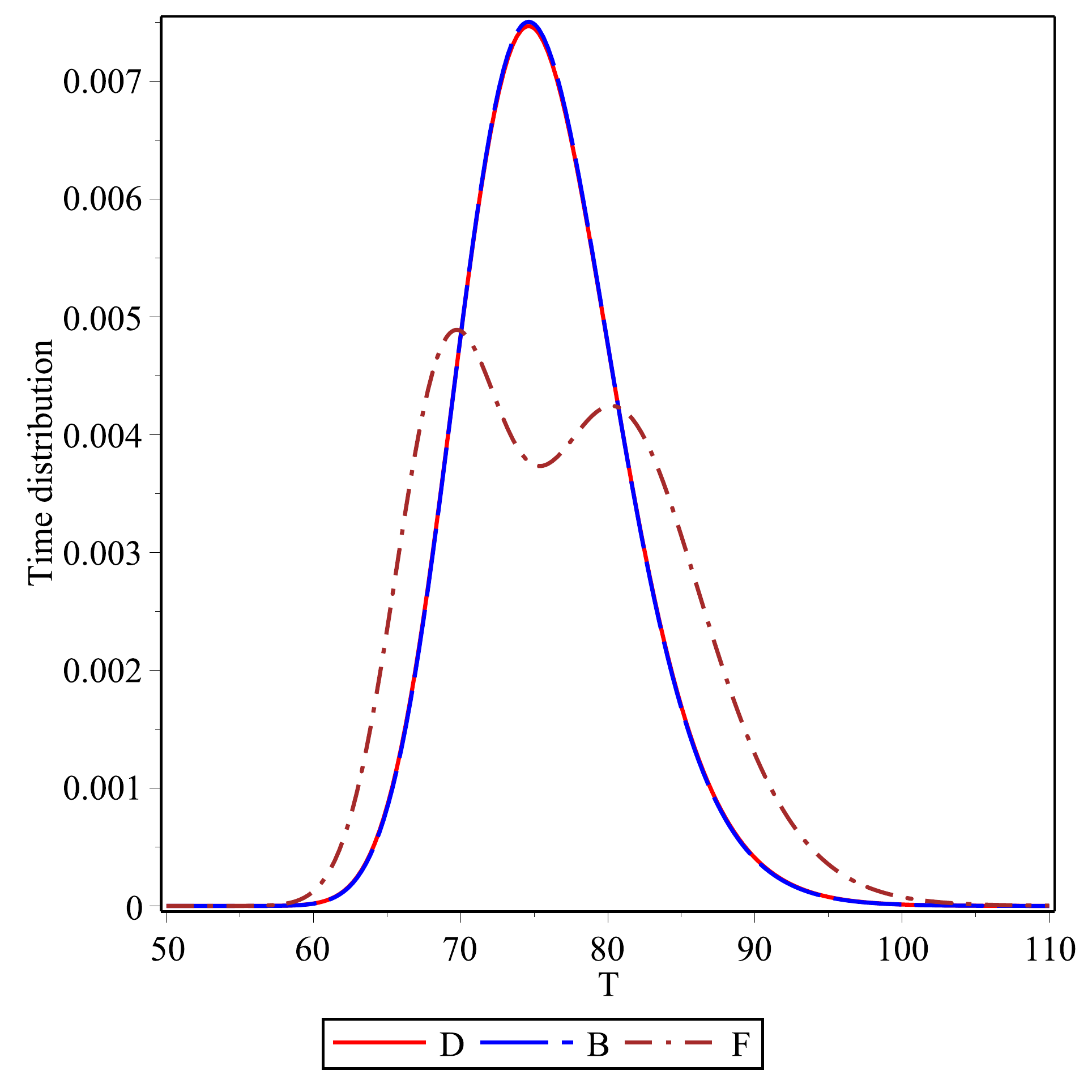}
	     \label{fig1d}
    \end{subfigure}
\caption{Non-relativistic free particle dynamics. The left panels show the initial spatial distribution, the right panels the relative one particle flight time distributions as defined in Eq. \ref{2.21} hitting the screen located at $X_f=450$. The other parameters used are $\Gamma= 0.01$ for the initial width of the coherent states; for case (I) (initial spatial difference) $X_{1i}=-301$, $X_{2i}= -299$, $K_{1i}=K_{2i}=10$, left and right top panels; and for case (II) (initial momentum difference) $X_{1i}=X_{2i}= -300$,  $K_{1i}=10.1$, $K_{2i}=9.9$, left and right bottom panels.} \label{fig1}
\end{figure} 

\begin{figure}
	\begin{subfigure}[t]{0.45\textwidth}
		\includegraphics[width=\textwidth]{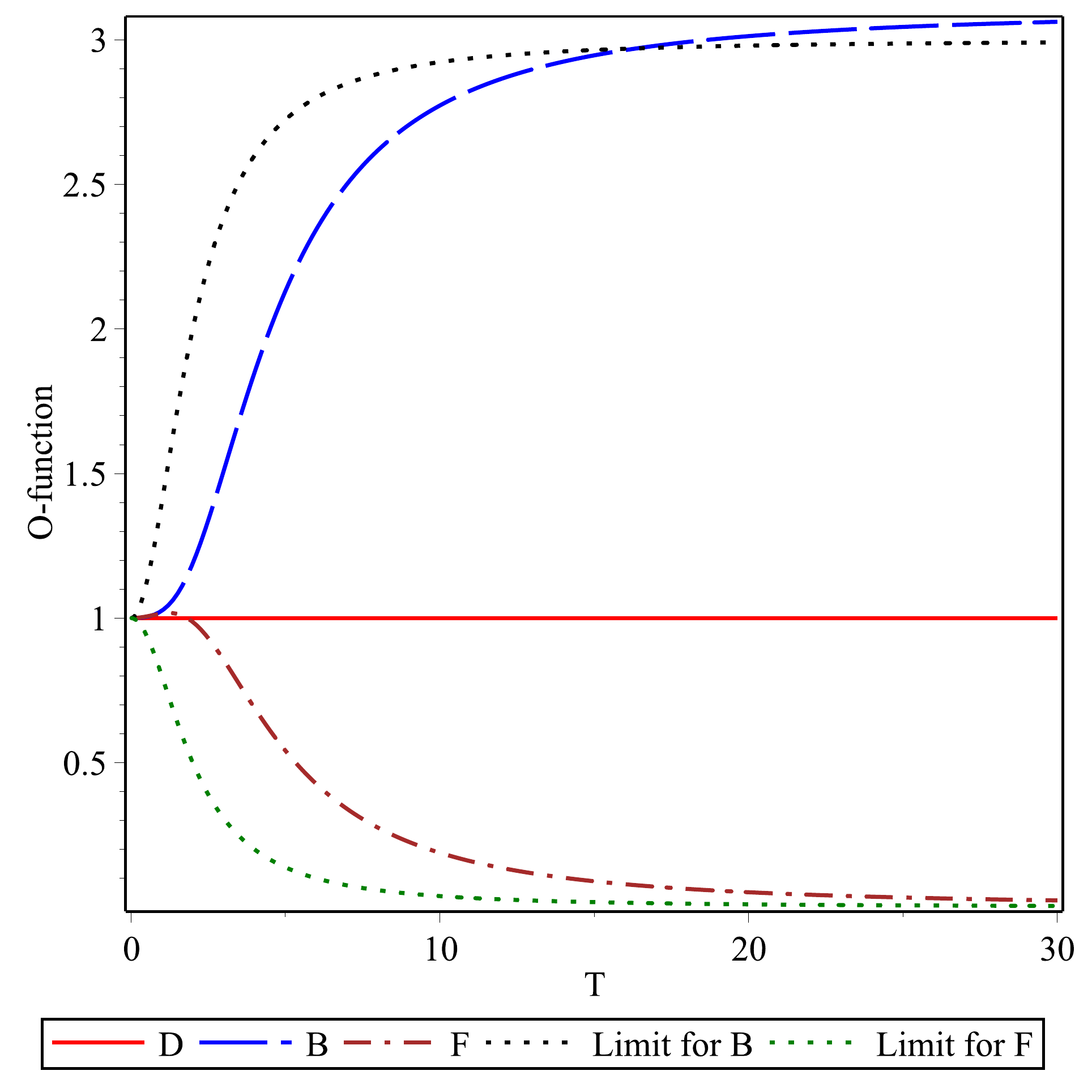}
		\label{fig2a}
	\end{subfigure}
	\hfill
	\begin{subfigure}[t]{0.45\textwidth}
		\includegraphics[width=\textwidth]{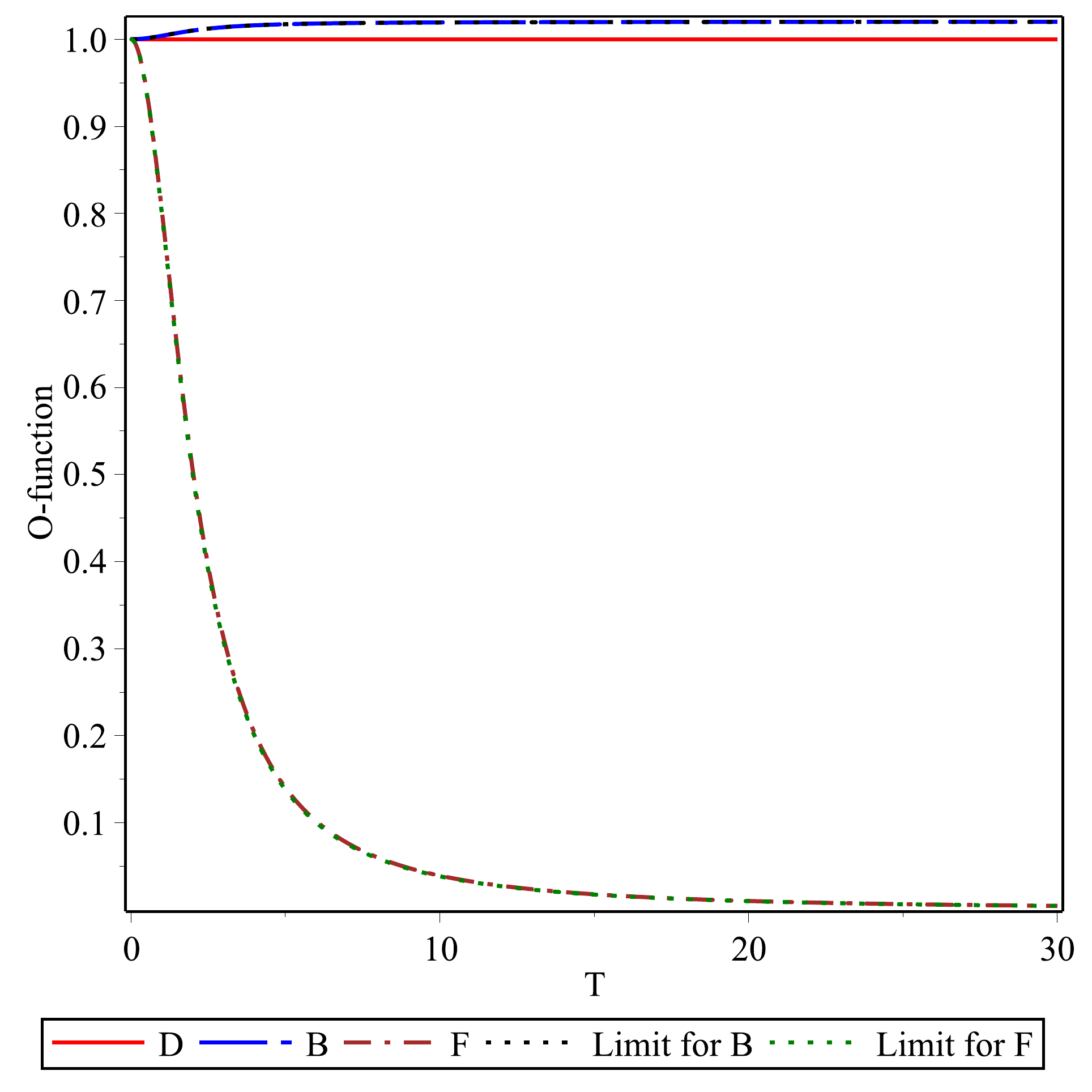}
		\label{fig2b}
	\end{subfigure}

	\caption{Overlap decay (O-function) defined in Eq. \ref{2.34} for non-relativistic free particles. The left and right panels are for initial spatial (case I) and momentum (case II) differences, respectively. The initial width of the Gaussians here is taken to be $\Gamma=0.01$. } \label{fig2}
\end{figure}

In Figure \ref{fig1}, we plot the initial spatial  distributions (left panels) and (see Eq. \ref{2.12}) the one-particle flight time distributions (right panels) for the two sets of initial conditions. The top two panels are for case (I) (initial spatial difference) the bottom two are for case  (II) (initial momentum difference).
The solid red curve is used for distinguishable particles,   the
long-dashed blue curve for bosons and the dot-dashed brown curve for fermions. The one-particle flight time distribution is broader for fermions displaying early and late arrivals. Bosons tend to arrive later than distinguishable particles, showing the narrowest time distribution.
Furthermore, the flight time distributions of distinguishable particles
and bosons tend to have a similar shape, losing the two lobes of the initial density, whereas 
the fermions display a double-peaked time distribution, reflecting their initial
density. In the bottom-right panel, bosons and distingusiable particles behave essentially identically
whereas fermions display a bimodal time distribution. Fermions not only arrive at the screen earlier and later, there is a distinctive time asymmetry 
in the flight time distribution.
We attribute these different behaviors to the bunching and anti-bunching properties of bosons and fermions, respectively. Specifically, the early arrival of fermions at the screen is related to the `front' of the initial fermionic density distribution, which, due to the anti-bunching effect, is closer to the origin than in the case of bosons and distinguishable particles. The same is true for the portion of the fermionic flight time distribution which arrives later at the barrier. It is due to the back of the initial fermionic density which is further from the origin, when compared to bosons or distinguishable particles.

In a previous work where we studied the MacColl-Hartman effect we have argued against the `front' of the wavepacket being used to explain away supposedly superluminal propagation for \textit{tunneling} particles \cite{dumont2020,rivlin2020}, noting that the superluminality cannot be used for the purpose of early signaling. When considering the MacColl-Hartman effect, one is comparing final time distributions of tunneled particles with free particles, but the two have the same initial density distribution. In the case considered here, the initial wavepacket of the fermions is broadened when compared to that of the bosons. It is this broadening which leads to early and late arrival times of fermions as compared to bosons, meaning that one does not have to consider here the possibility of superluminality.

Another interesting aspect considered here is 
the analysis of the survival amplitude, using Eq. (\ref{2.34}) 
and the limits at small initial spatial $\Delta_X$ and momentum $\Delta_K$ difference and the associated time dependence, as in Eqs. (\ref{2.38}) and (\ref{2.39}). In Figure \ref{fig2} we plot the overlapping functions (O-function, see Eq. \ref{2.34}) for  initial conditions (I) (initial spatial difference) in the left panel and for case (II) (initial momentum difference) in the right panel. 
As in Fig. \ref{fig1}, the red line indicates the behavior for distinguishable particles, the long-dashed blue curve is used for bosons and the dot-dashed brown curve for 
fermions. Dotted black and green curves correspond to the limits of small values of  $\Delta_X$, $\Delta_K$ and time. The time dependence of the corresponding overlapping functions is quite different for bosons and fermions. The bosons stay together for a long time whereas the fermionic overlap decays rapidly.

\subsection{Free particle relativistic flight time distributions}

Figure 3 shows the time-dependent density at the screen (at $X=0$) for two photons and two electrons traveling near the speed of light.  In the left panel, the two Gaussians are centered at $X_{1i}=-3.5$ and $X_{2i}=-3$ and at a wavenumber consistent with velocity, $v=0.99c$.  In the right panel, the two Gaussians are centered at $X_{1i}=X_{2i}=-3$ and at wavenumbers consistent with $v=0.984c$ and $0.996c$.  In both cases, an electron is more likely to arrive at the screen before a photon.  However, this is simply because the initial density for the electrons is broader than that of the photons.  To show this, the density that would be seen if the electrons traveled dispersion-free at the speed of light is also shown (dotted lines).  The observed electron density clearly travels with a speed less than $c$. The early and late arrivals of the fermions is just a reflection of their initial density, which, as may be seen from the left panels of Fig. \ref{fig1}, is broader than the initial distribution for bosons, due to the anti-bunching effect of fermions. The early arrival times are a reflection of the initial width of the packet -- this is the same for both the relativistic and the non-relativistic regimes.

\begin{figure}
	\begin{subfigure}[t]{0.45\textwidth}
		\includegraphics[width=\textwidth]{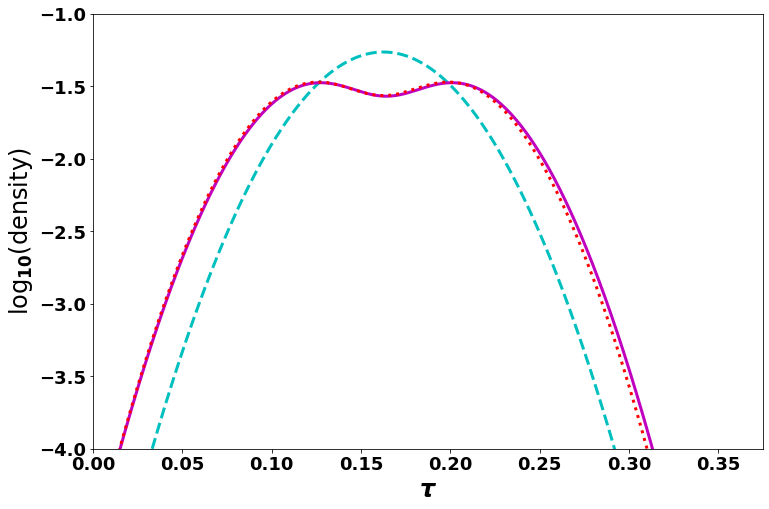}
		\label{fig3a}
	\end{subfigure}
	\hfill
	\begin{subfigure}[t]{0.45\textwidth}
		\includegraphics[width=\textwidth]{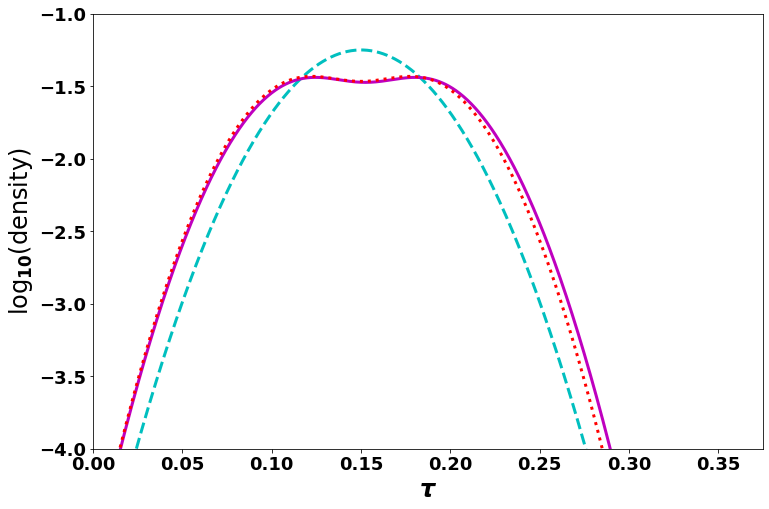}
		\label{fig3b}
	\end{subfigure}

\caption{\noindent Time-dependent densities for two photons (dashed lines) and two relativistic electrons (solid lines). The left and right panels are for initial spatial (case I) and momentum (case II) differences, respectively. The initial width of the Gaussians here is taken to be $\Gamma=0.0025$. Also shown (dotted lines) are the densities that would be seen if the electrons traveled dispersion-free at the speed of light.} \label{fig3}
\end{figure}

\subsection{Identical particle, non-relativistic flight times for scattering on a delta potential barrier} 

In this case, as noted above, the mean flight times (Eq. \ref{2.10}) are well defined, and they provide a clear measure of the effect of particle symmetry on the flight time. 
In Table \ref{table1}, we provide the transmitted and reflected mean flight times (in reduced coordinates) for bosons $<\tau_{1,B}>_{T,R}$ and 
fermions $<\tau_{1,F}>_{T,R}$ with $\Gamma=0.01$ and $\epsilon=1$ for the two cases of initial spatial (I) and momentum (II) differences. Transmitted mean times 
are always shorter than reflected mean times due to the momentum filtering effect and those of  fermions are always greater than for bosons due to their anti-bunching
and bunching properties. 

\begin{table}
	\caption{Initial conditions  and transmitted (T) and reflected (R) mean flight times (in reduced coordinates) for bosons, $<\tau_{1,B}>_{T,R}$, and fermions, $<\tau_{1,F}>_{T,R}$, with $\Gamma=0.01$ and $\varepsilon=1$.} 
	\begin{tabular}{||c|c|c|c||c|c||c|c||}
		\hline       $X_{1i}$& $X_{2i}$ & $X_f$ & $K_{1i}, K_{2i}$  & $<\tau_{1,B}>_T$ &  $<\tau_{1,B}>_R $  & $<\tau_{1,F}>_T$ &  $<\tau_{1,F}>_R $ \\
		\hline    -301 &  -299 & $\pm$ 450 & 10, 10 &75.2908 & 75.8847  & 75.4992  & 76.5237 \\
		\hline     -300 &  -300& $\pm$ 450 & 10.1, 9.9 &75.3749 & 76.1740 & 76.7417  &77.3525  \\
		\hline
		\hline 
	\end{tabular}
	\label{table1}
\end{table}

\begin{figure}
	\begin{subfigure}[t]{0.45\textwidth}
		\includegraphics[width=\textwidth]{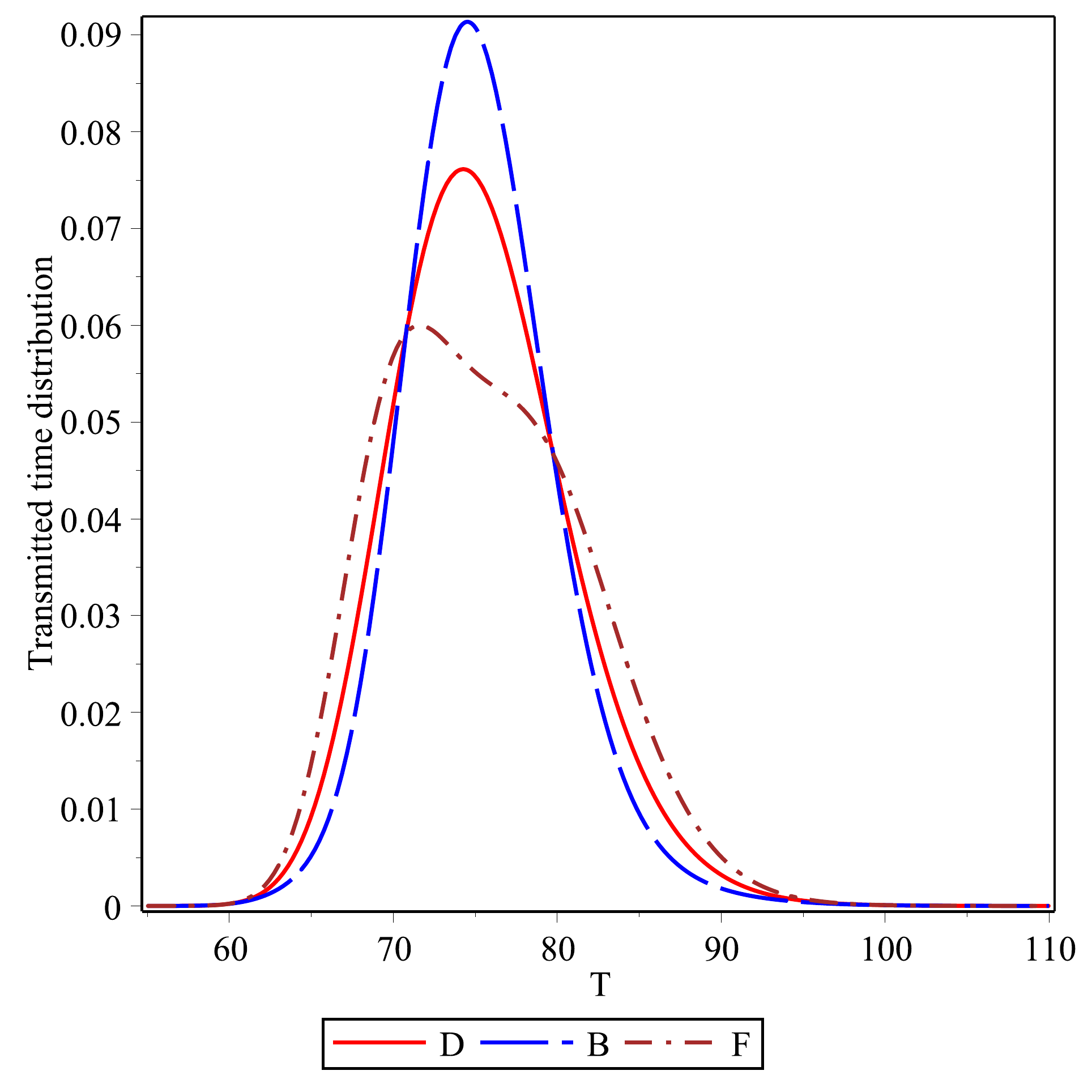}
		\label{fig1a}
	\end{subfigure}
	\hfill
	\begin{subfigure}[t]{0.45\textwidth}
		\includegraphics[width=\textwidth]{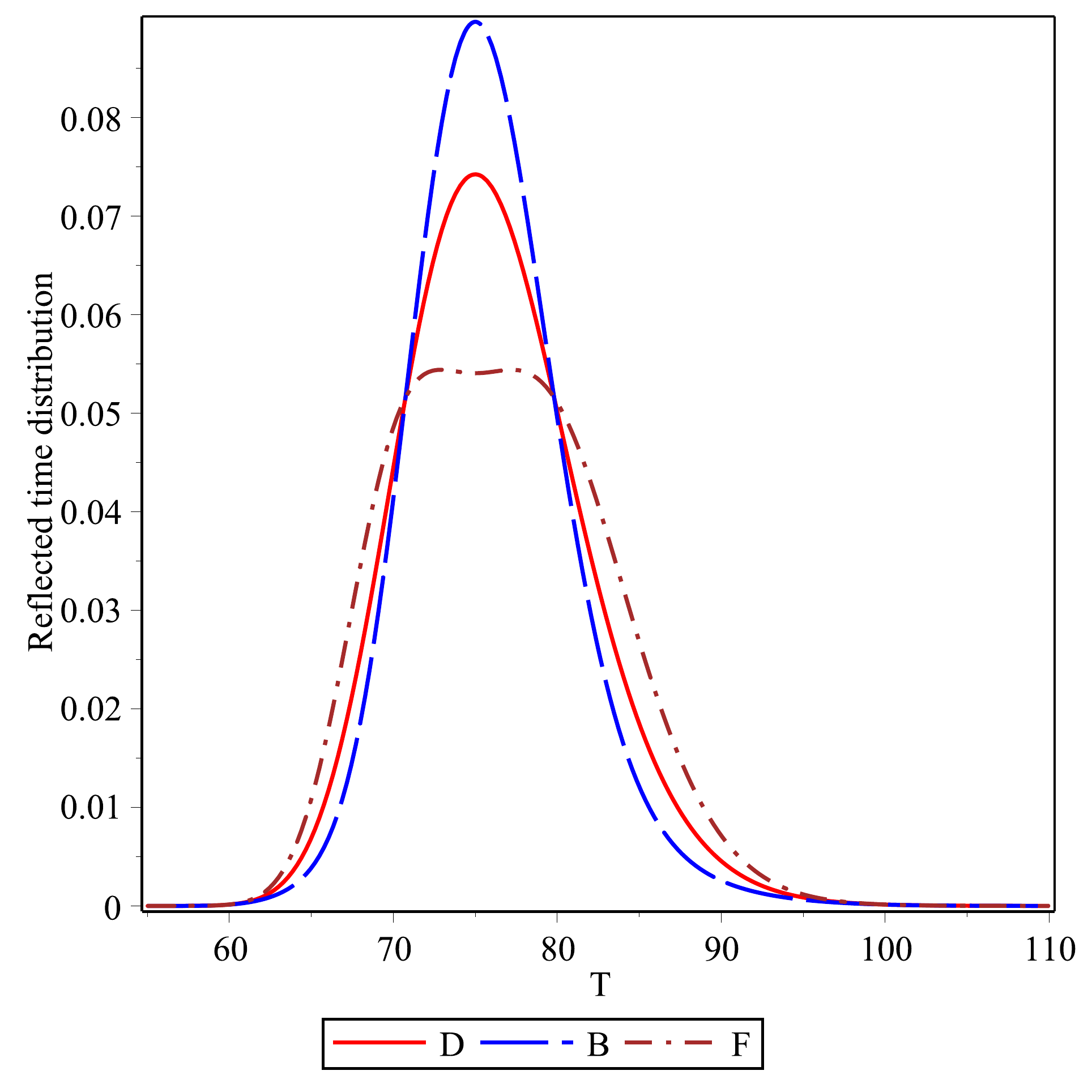}
		\label{fig1b}
	\end{subfigure}
	
	\begin{subfigure}[b]{0.45\textwidth}
		\includegraphics[width=\textwidth]{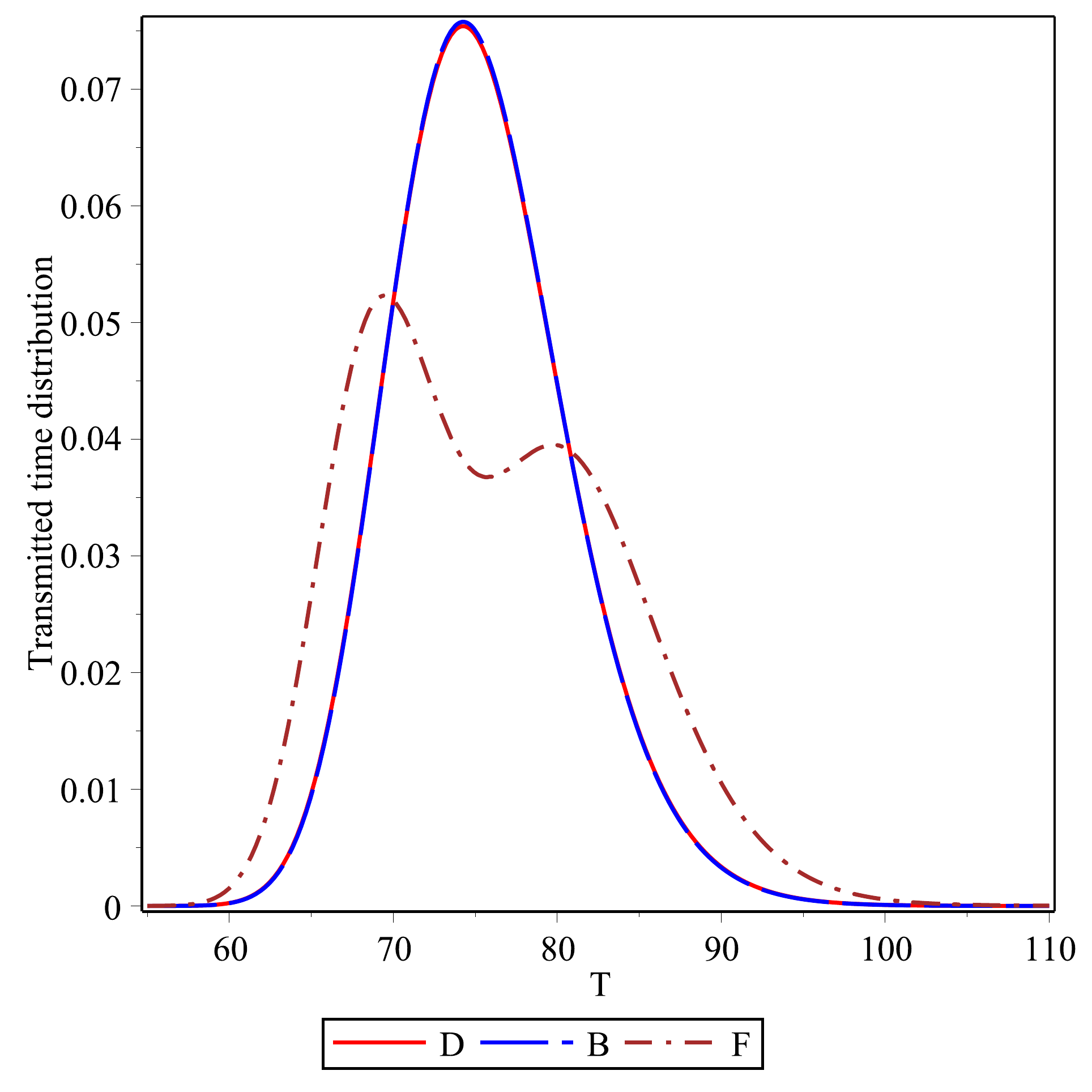}
		\label{fig1c}
	\end{subfigure}
	\hfill
	\begin{subfigure}[b]{0.45\textwidth}
		\includegraphics[width=\textwidth]{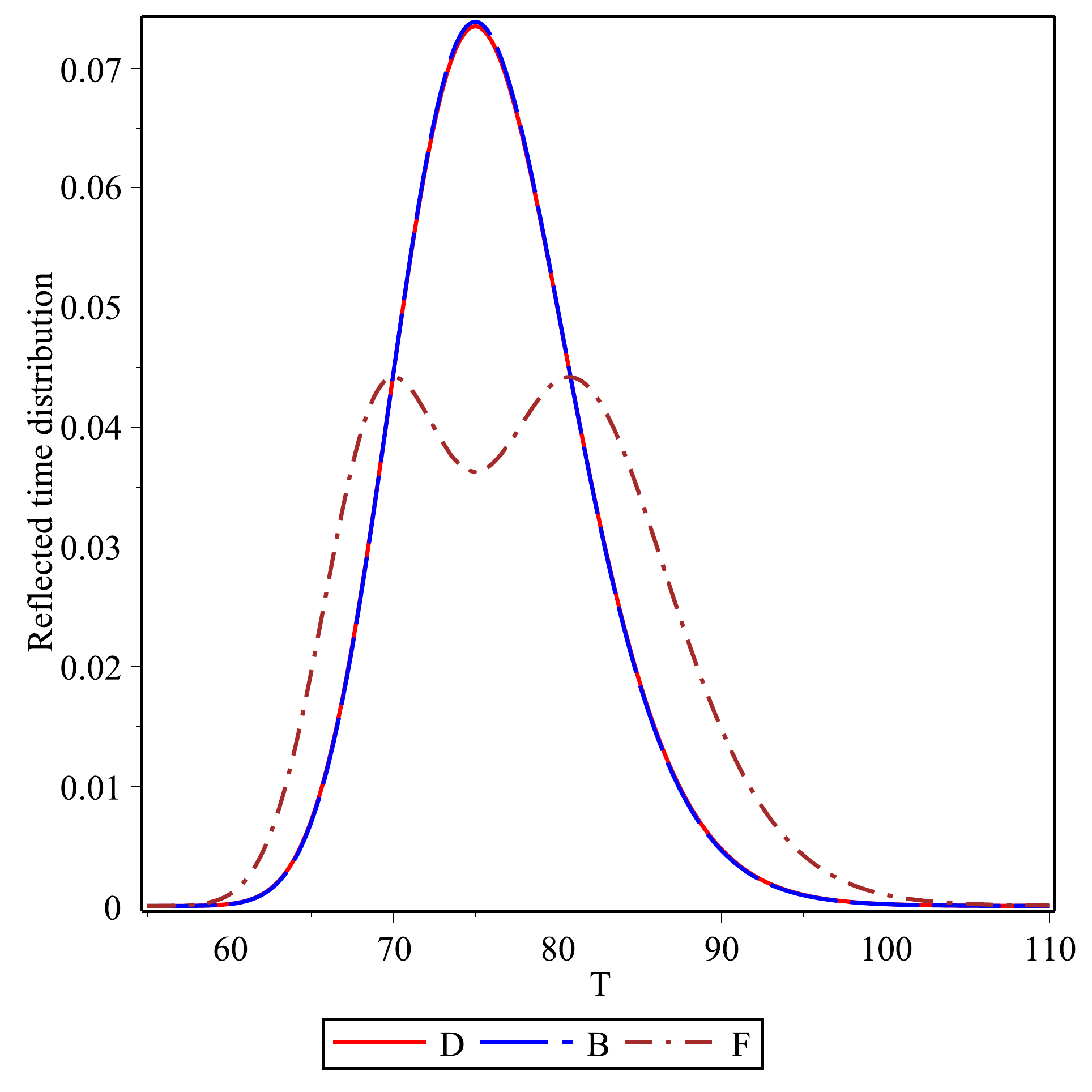}
		\label{fig1d}
	\end{subfigure}
	\caption{Tunneling $\delta$-barrier dynamics with $\Gamma= 0.01$. One-particle transmitted and reflected mean flight time distributions  for  conditions (I) are shown in the  left and right top panels and  for conditions (II)  in the left and right bottom panels.  } \label{fig4}
\end{figure}

The corresponding flight time probability distributions (eq. \ref{2.11}) are shown in Fig. \ref{fig4}.  
The top and bottom panels correspond to initial spatial (case I) and initial momentum (case II) differences, the left and right panels correspond to the transmitted and reflected flight time distributions, respectively. The trends are similar to those found in the free particle dynamics scenario. The effect of symmetry on flight times seems to be robust and is not changed much in the presence of an interaction barrier. Time distributions are broadest for fermions and narrowest for bosons, with distinguishable particles in between. The asymmetry of the bimodal reflected distributions for fermions becomes less important when comparing with the transmitted ones.

Finally, it is also of interest to analyze the role played by the initial width ($\Gamma$) of the Gaussian wavepackets in the mean tunneling flight times which are well defined, especially when $\Gamma$ approaches zero such that the spatial extent of the two Gaussians is large, which creates large initial overlaps. This analysis is carried out by means of a fitting procedure to a linear function ($b \Gamma + c$) obtained from the numerics for eight values of the initial width,
$\Gamma= 10^{-2}, 0.25 \, 10^{-2}, 9.0 \, 10^{-4}, 4.0 \, 10^{-4}, 10^{-4}, 0.25 \, 10^{-4}, 9.0 \, 10^{-6}, 4.0 \, 10^{-6}$.
In Figure \ref{fig3}, the transmitted  (left panel) and reflected (right panel) mean flight times versus $\Gamma$ for distinguishable particles, bosons and fermions
are plotted.  The legend is the same one used along this work for each particle. In every case, the quality of the fitting is very good. All particles tend to a value of  $c= 75.005$, which is the phase time under conditions (I). Thus, the symmetry seems to play no role in the mean tunneling flight times since the phase time for a single particle is recovered in the limit $\Gamma \rightarrow 0$. 

\begin{figure}
	\begin{subfigure}[t]{0.45\textwidth}
		\includegraphics[width=\textwidth]{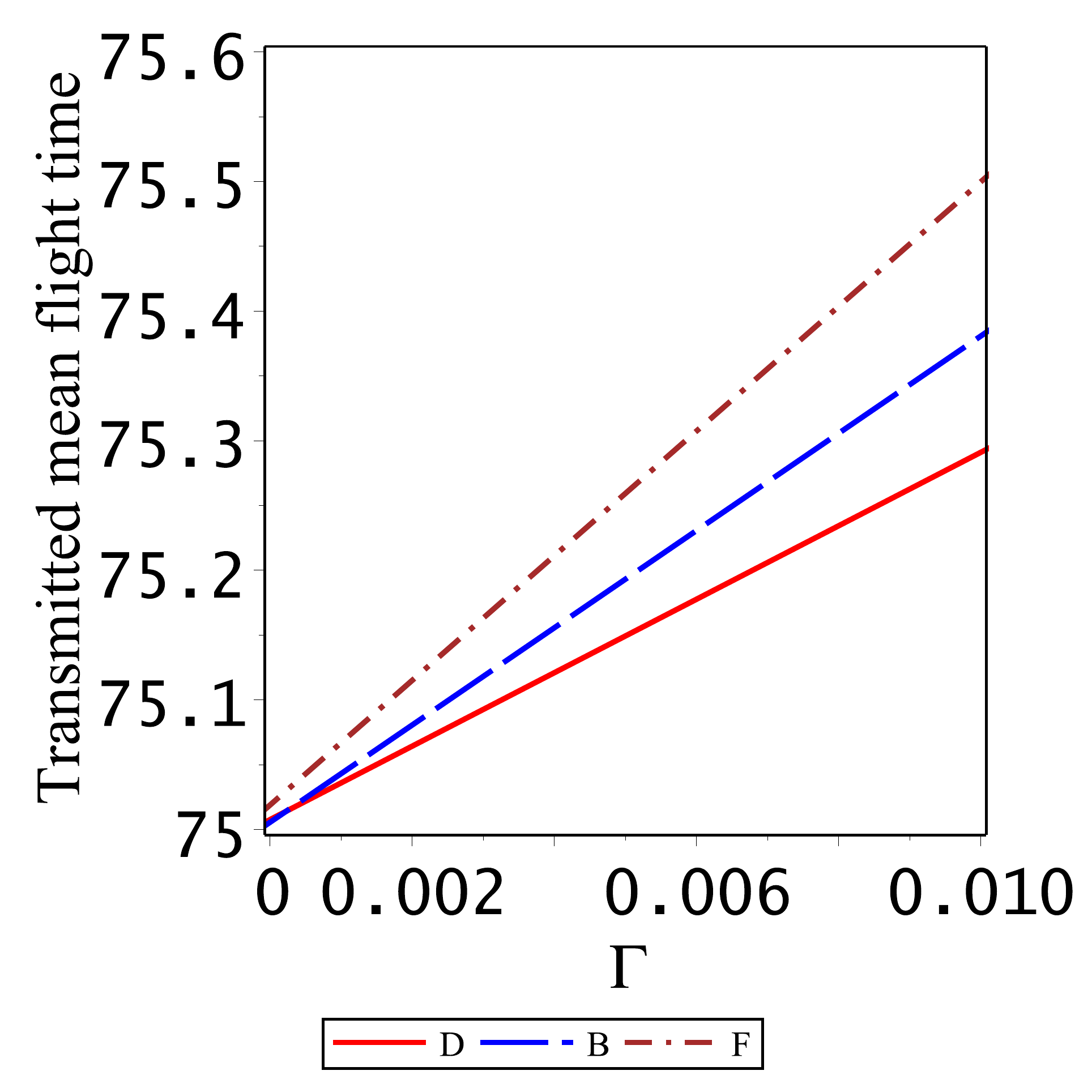}
		\label{fig2a}
	\end{subfigure}
	\hfill
	\begin{subfigure}[t]{0.45\textwidth}
		\includegraphics[width=\textwidth]{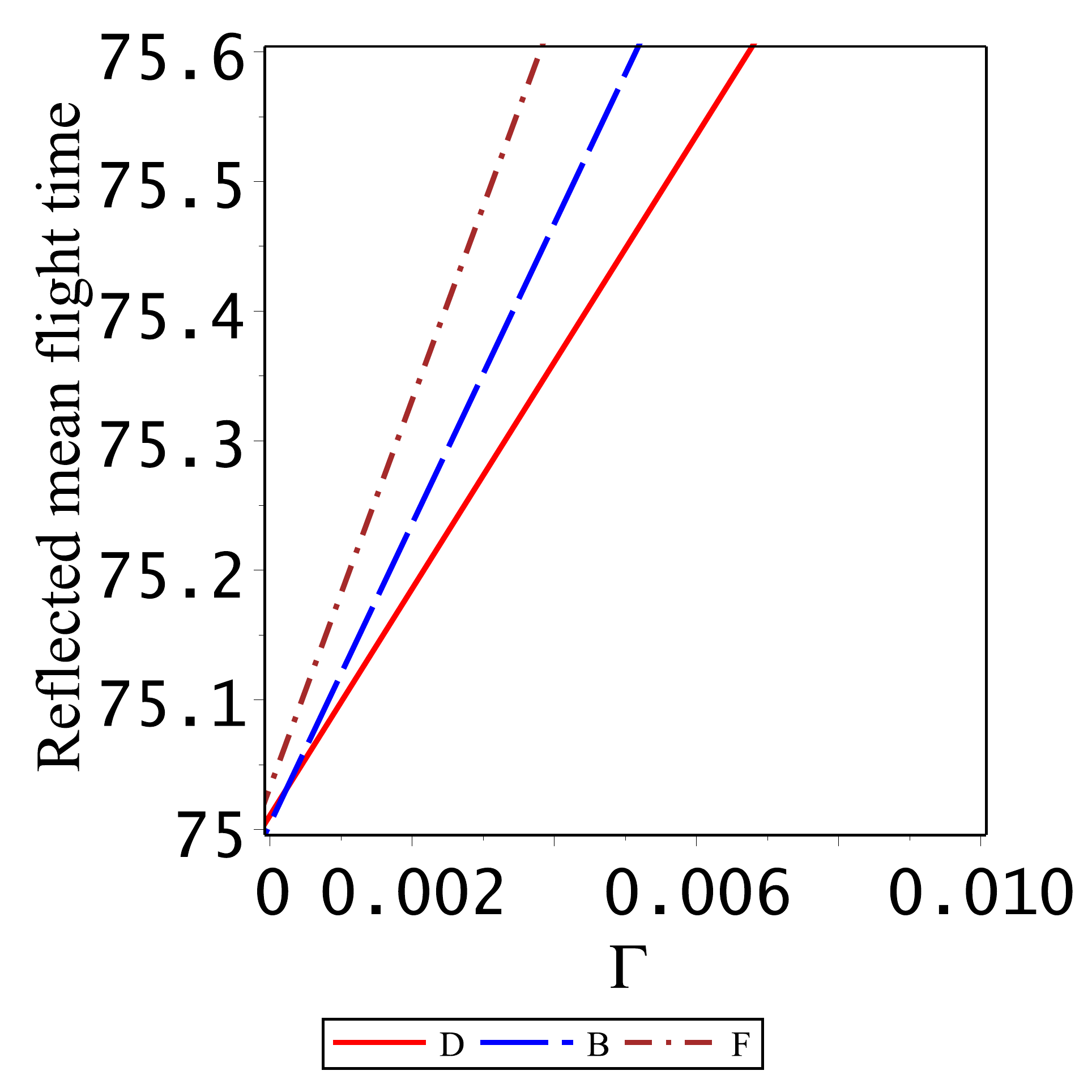}
		\label{fig2b}
	\end{subfigure}

	\caption{One-particle transmitted  (left panel) and reflected (right panel)  mean flight times versus $\Gamma$ for distinguishable particles, bosons and fermions under conditions (I). The legend is the same one used along this work for each particle.} \label{fig3}
\end{figure}

\section{Discussion}

In this work, we have analyzed the influence of the symmetry of the wavefunction for a system consisting of two non-interacting identical particles. The well-known bunching and anti-bunching properties exhibited by bosons and fermions respectively have been translated into the spreading of the initial spatial densities and flight time distributions under the presence or not of an interaction potential such as a delta barrier, leading to early arrivals for fermions. Interestingly, the symmetry of the wave function seems to be robust for flight time distributions, but it does not affect the mean tunneling flight times. Analyzing the short-time dynamics through the survival probability, fermions tend to decay faster than bosons, which has profound implications for the well-known Zeno effect.

In our model, we have considered non-interacting identical particles. This is clearly an oversimplification of the real problem since, for example, electrons interact with each other through the long-range Coulomb repulsion. As far as we know, very few studies have addressed the issue of the effect of symmetry on flight times and the Zeno effect even though it should be readily accessible. For example, the triplet states of two electrons will give a fermionic spatial wavefunction while the singlet state a bosonic one. In a scattering experiment similar to the one discussed in Refs. \cite{grossmann2014,buchholz2018}, the two particles in the center of mass frame will approach each other such that the distances between the two particles becomes small and the symmetry will affect the time of flight distribution of the two particles. Due to the anti-bunching property of fermions, one should expect broader temporal distributions of the scattered particles.  

Lozovik {\it et al} \cite{Filinov2} considered tunneling of two interacting particles in a double-well potential. They used quantum molecular dynamics within the Wigner representation and found that exchange effects are very important and affect the tunneling. However, this work does not mention flight time distributions. It is thus possible, at least in principle, to study the effect of particle symmetry on flight time distributions of identical electrons, without neglecting the repulsive potential of interaction between them, though the actual numerical implementation, especially in the relativistic regime is much more challenging. 

Finally, we note that  the study of symmetry on flight time distributions presented in this paper may be generalized to anyons, by introducing a phase in the initial distribution.

\vspace{1cm}
\noindent
{\bf Acknowledgements}
TR and EP acknowledge support from the Israel Science Foundation, SMA acknowledges support from the Ministerio de Ciencia, Innovaci\'on y Universidades (Spain) 
under the Project FIS2017-83473-C2-1-P and Fundaci\'on Humanismo y Ciencia.

\end{document}